\documentclass[letterpaper,twocolumn,10pt]{article}
\usepackage{usenix2019_v3}
\usepackage[official]{eurosym}

\usepackage{booktabs}

\usepackage{array}
\newcolumntype{C}[1]{>{\centering\let\newline\\\arraybackslash\hspace{0pt}}m{#1}}
\newcolumntype{L}[1]{>{\raggedright\let\newline\\\arraybackslash\hspace{0pt}}m{#1}}
\newcolumntype{R}[1]{>{\raggedleft\let\newline\\\arraybackslash\hspace{0pt}}m{#1}}

\usepackage{hyperref}
\usepackage[inline]{enumitem}
\usepackage[]{xcolor}
\usepackage{threeparttable}
\usepackage{amsmath,multirow,array}
\usepackage{booktabs}
\usepackage{ifthen}
\usepackage[numbers,sort]{natbib}
\usepackage{subcaption}

\usepackage{tikz}
\usetikzlibrary{decorations.pathmorphing}
\usetikzlibrary{decorations.markings}
\usetikzlibrary{decorations.pathreplacing}
\usetikzlibrary{calc}
\usetikzlibrary{backgrounds}
\usetikzlibrary{shapes}
\tikzset{>=stealth}

\definecolor{red_c}{rgb}{0.89,0.10,0.11}
\definecolor{blue_c}{rgb}{0.15,0.35,0.50}
\definecolor{green_c}{rgb}{0.24,0.54,0.23}
\definecolor{yellow_c}{rgb}{1, 0.769, 0.145}

\definecolor{darkpurple}{HTML}{C51B7D}
\definecolor{mediumpurple}{HTML}{DE77AE}
\definecolor{mediumgreen}{HTML}{7FBC41}
\definecolor{darkgreen}{HTML}{4D9221}
\definecolor{plbtgreen}{cmyk}{0.14,0,0.70,0.25}
\definecolor{maincolor}{cmyk}{0,0.5,1,0} 
\definecolor{altcolor}{cmyk}{1,0.6,0,0.56} 

\definecolor{p1color}{HTML}{E16A86}
\definecolor{p2color}{HTML}{AA9000}
\definecolor{p3color}{HTML}{00AA5A}
\definecolor{p4color}{HTML}{00A6CA}
\definecolor{p5color}{HTML}{B675E0}

\definecolor{IdkColor}{HTML}{1B1B1B}
\colorlet{IdkColor}{IdkColor}
\colorlet{VeryNegativeColor}{red_c}
\colorlet{NegativeColor}{red_c!20}
\colorlet{NeutralColor}{altcolor!60}

\colorlet{PositiveColor}{green_c!20}
\colorlet{VeryPositiveColor}{green_c}

\newcommand{\Aq}{\vphantom{Aq}}
\newcolumntype{C}[1]{>{\centering\let\newline\\\arraybackslash\hspace{0pt}}m{#1}}
\newcolumntype{L}[1]{>{\raggedright\let\newline\\\arraybackslash\hspace{0pt}}m{#1}}
\newcolumntype{R}[1]{>{\raggedleft\let\newline\\\arraybackslash\hspace{0pt}}m{#1}}
\newcommand{\ordinalthree}[4]{
	\tikz[x=.15mm,yshift =1.5ex]{\draw[line width=2ex, color = #4!20] (0,0)--node[black]{\ifthenelse{\equal{#1}{0} \OR \equal{#1}{1} \OR \equal{#1}{2} \OR \equal{#1}{3} \OR \equal{#1}{4}}{}{\tiny#1}}(#1,0);
		\draw[line width=2ex, color = #4!60] (#1,0)--node[black]{\ifthenelse{\equal{#2}{3} \OR \equal{#2}{4}}{}{\tiny#2}}++(#2,0);
		\draw[line width=2ex, color = #4] (#1,0)++(#2,0)--node[white]{\ifthenelse{\equal{#3}{0} \OR  \equal{#3}{1} \OR \equal{#3}{2} \OR \equal{#3}{3} \OR \equal{#3}{4}}{}{\tiny#3}}(100,0);
	}
}

\newcommand{\ordinalthreerev}[4]{
	\tikz[x=.15mm,yshift =1.5ex]{\draw[line width=2ex, color = #4] (0,0)--node[black]{\ifthenelse{\equal{#1}{0} \OR \equal{#1}{1} \OR \equal{#1}{2} \OR \equal{#1}{3} \OR \equal{#1}{4}}{}{\tiny#1}}(#1,0);
		\draw[line width=2ex, color = #4!60] (#1,0)--node[black]{\ifthenelse{\equal{#2}{3} \OR \equal{#2}{4}}{}{\tiny#2}}++(#2,0);
		\draw[line width=2ex, color = #4!20] (#1,0)++(#2,0)--node[white]{\ifthenelse{\equal{#3}{0} \OR  \equal{#3}{1} \OR \equal{#3}{2} \OR \equal{#3}{3} \OR \equal{#3}{4}}{}{\tiny#3}}(100,0);
	}
}

\newcommand{\ordinalfour}[5]{
	\tikz[x=.15mm,yshift =1.5ex]{\draw[line width=2ex, color = #5!20] (0,0)--node[black]{\ifthenelse{\equal{#1}{0} \OR \equal{#1}{1} \OR \equal{#1}{2} \OR \equal{#1}{3} \OR \equal{#1}{4}}{}{\tiny#1}}(#1,0);
		\draw[line width=2ex, color = #5!60] (#1,0)--node[black]{\ifthenelse{\equal{#2}{3} \OR \equal{#2}{4}}{}{\tiny#2}}++(#2,0);
		\draw[line width=2ex, color = #5] (#1,0)++(#2,0)--node[white]{\ifthenelse{\equal{#3}{0} \OR  \equal{#3}{1} \OR \equal{#3}{2} \OR \equal{#3}{3} \OR \equal{#3}{4}}{}{\tiny#3}}++(#3,0);
		\draw[line width=2ex, color = yellow!50] (#1,0)++(#2,0)++(#3,0)--node[white]{\ifthenelse{\equal{#4}{0} \OR  \equal{#4}{1} \OR \equal{#4}{2} \OR \equal{#4}{3} \OR \equal{#4}{4}}{}{\tiny#4}}(100,0);
	}
}
\def\anonymised{1} 

\def\questionnaire{1} 

\def\true{1}

\newcommand{\ifanonym}[2]{\ifx\anonymised\true#1\else#2\fi}
\newcommand{\ifquestionnaire}[2]{\ifx\questionnaire\true#1\else#2\fi}
\newcommand{\dummylink}{[\emph{link to the static webpage}]}
\newcommand{\dummyresearcher}{[\emph{first author's name}]}
\newcommand{\dummyuni}{[\emph{researchers' affiliation}]}
\newcommand{\dummycontact}{[\emph{contact e-mail address}]}
\newcommand{\dummydataprotec}{[\emph{DPO e-mail address}]}

\newcommand{\questiontype}[1]{\strut\hfill{#1}}
\newcommand{\singlechoice}{\questiontype{(Single~choice)}}
\newcommand{\multiplechoice}{\questiontype{(Multiple~choice)}}
\newcommand{\opentext}{\questiontype{(Open~text)}}

\definecolor{yesbefore}{rgb}{0.35,0.7,0.9}
\definecolor{yesafter}{rgb}{0,0.45,0.7}
\definecolor{no}{rgb}{0.80,0.40,0}

\usepackage{xurl}
\usetikzlibrary{calc,shadows,decorations,shapes.symbols}

\begin{document}

\date{}

\title{\Large \bf 

	Anatomy of a High-Profile Data Breach:\\
	Dissecting the Aftermath of a Crypto-Wallet Case  }

\author{
{\rm Svetlana Abramova}\\
Universität Innsbruck
\and
{\rm Rainer Böhme}\\
Universität Innsbruck
} 

\maketitle

\begin{abstract}
Media reports show an alarming increase of data breaches at providers of cybersecurity products and services. 
Since the exposed records may reveal security-relevant data, such incidents cause undue burden and create the risk of re-victimization to individuals whose personal data gets exposed. 
In pursuit of examining a broad spectrum of the downstream effects on victims, we surveyed 104 persons who purchased specialized devices for the secure storage of crypto-assets and later fell victim to a breach of customer data. 
Our case study reveals common nuisances (i.\,e., spam, scams, phishing e-mails) as well as previously unseen attack vectors (e.\,g., involving tampered devices), which are possibly tied to the breach. 
A few victims report losses of digital assets as a form of the harm. 
We find that our participants exhibit heightened safety concerns, appear skeptical about litigation efforts, and demonstrate the ability to differentiate between the quality of the security product and the circumstances of the breach. 
We derive implications for the cybersecurity industry at large, and point out methodological challenges in data breach research.
\end{abstract}
\section{Introduction}
Data breaches (i.\,e., the leakage or disclosure of sensitive, confidential, or otherwise protected information to unauthorized parties~\cite{romanosky2011data}) continue to plague businesses and customers around the globe. 
As these unwanted events become commonplace in some industries~\cite{wolff2016ex}, it is worrying that more and more cybersecurity vendors and service providers appear in the headlines of reports announcing new leaks. 
Last year's notorious examples include Microsoft, with over two terabytes of disclosed business customer data~\cite{Microsoft2022}, and the LastPass password manager, which suffered from an unauthorized access to its backup of customer vault data~\cite{LastPass2022}.

Breach cases of this kind are noteworthy for two reasons. 
First, they demonstrate that even the cybersecurity industry cannot effectively prevent breaches and thus must adopt a \emph{`not if, but when'} security mindset. 
Security providers must prepare for potential breach events, so that their crisis communication can respond in a fully transparent, timely and instructive manner. 
Second, high-profile breaches may place a particularly high burden on the individuals whose personal data gets exposed, because these events can serve as stepping stones for future attacks. 
Customer data leaked by security providers may inadvertently supply threat actors with valuable facts about people and their security actions. 
As a result, victims become subject to increased and more targeted attacks, which force them to maintain high security vigilance. 

To date, data breach research has not studied this kind of incidents and wider implications thereof. 
In fact, there is still a lack of knowledge and methods to estimate the downstream effects of conventional (i.\,e., lower-profile) breaches on customers and their post-breach behavior. 
Only recently, scholars began to study victims' reactions, such as attitude changes concerning risk and trust~\cite{bansal2021you,ayaburi2020effect,zou2018ve}, emotional~\cite{bachura2022opm,novak2019internet,ivaturi2020mapping} and behavioral responses~\cite{mayer2021now,liang2020customer,bhagavatula2020people,turjeman2019data,zou2018ve}, monetary and psychological harms~\cite{kilovaty2021psychological}, and perceptions of litigation actions~\cite{G2021322}. 
However, many studies either narrow their scope of analysis to a single type of an attack or behavior, or discuss hypothetical scenarios, for which customer-facing consequences are overly general or unfounded. 
Effects may take many forms and, therefore, neither approach yields an in-depth picture of the aftermath of a breach.

To address these research gaps, we present a case study of the data breach affecting Ledger customers, which happened in July 2020~\cite{Ledger2020}. 
This incident meets our criteria of being `specific', because it affected one of the leading manufacturers of hardware wallet devices intended for secure offline storage of crypto-assets. 
Personal data (including names, postal and e-mail addresses, phone numbers) of a subset of the company's global customer base got publicly exposed, while the purchased security devices were not  compromised.

We are interested in examining the effects of this particular event for several reasons. 
Crypto-asset owners are attractive targets for both amateur and professional fraudsters looking for  ways to monetize the leaked data. 
Compared to the average internet user, crypto-asset owners who proactively prefer specialized storage devices demonstrate higher security awareness and caution~\cite{AVBB2021-CHI,lindqvist2021bitcoin}. 
Hence, they might be less susceptible to successful crimes and serious damages. 
Finally, security-aware victims of this incident may serve as a reliable and knowledgeable source of information when it comes to reporting about the aftermath of the breach. 

Our study focuses on the three key research questions covering a broad spectrum of the potential effects of a data breach. 
First, we aim to elicit \emph{which harms}, be they financial, emotional distress, or invasion of privacy, have been experienced, and in \emph{what ways} this incident has impacted the affected persons (RQ1). 
We are guided by insights from prior research suggesting that leaked sensitive data results in an increased risk of online identity theft~\cite{romanosky2014empirical}, account compromise~\cite{thomas2017data}, and phishing~\cite{peng2019happens}.
Furthermore, cryptocurrency markets themselves are deemed to be prone to fraud, thefts, and hacking~\cite{bartoletti2021cryptocurrency,HAMRICK2021102506}. 
This suggests potentially heightened interest in the leaked data from perpetrators specializing in this domain. 
By collecting empirical evidence of the experienced threats and harms, we set to build a knowledge base, sought-after also in policy debates on  data breach legislation~\cite{romanosky2011data,kilovaty2021psychological}.

Concerning our second research question: in order to mitigate harm, victims are advised to adopt protective measures, such as changing passwords~\cite{fagan2016they}, adding two-factor authentication~\cite{colnago2018s}, and managing multiple e-mail aliases for account registration~\cite{mayer2021now}. 
However, these advisable actions are rarely observable in practice, as past studies repeatedly show~\cite{mayer2021now,zouCHI2020}. 
Drawing on these results, we seek to investigate which responses and changes in individuals' security behaviors this breach has triggered (RQ2). 
The fact that crypto-assets (without consumer protection and little chance of legal recourse) are at stake, adds relevance to this research question. 

Third, the analysis of a breach affecting a security provider would be incomplete without studying far-reaching impacts on customer attitudes, trust, and loyalty. 
To this end, we sought inspiration in the marketing literature~\cite{gurhan2004corporate,brown1997company}, which suggests separating an individual's corporate and product associations. 
With RQ3, we intend to explore to what extent the breach may have affected consumer attitudes to the company itself as compared with the impact on its security product.

To address these questions, we recruited 104 participants using e-mail addresses from the leaked dataset. 
This recruitment method makes us confident that many of our respondents are in fact victims of the breach. 
By implementing a series of risk mitigation measures outlined below in the paper, our study was approved by the Institutional Review Board (IRB). 

The rest of this paper is organized as follows. 
We first present our case, including a timeline of relevant events  (Section~\ref{section:case}). 
Then, we review the state-of-the-art literature on data breaches (Section~\ref{section:relwork}). 
Next, we thoroughly explain the methodology of our study (Section~\ref{section:method}). 
We present the empirical results broken down by research question in Section~\ref{section:results}. 
Finally, we discuss the implications of our results for practitioners and researchers, recall limitations, and conclude (Sections~\ref{section:discussion}, \ref{section:conclusion}). 
\section{The Case}\label{section:case}
Ledger,\footnote{\url{https://www.ledger.com/}} headquartered in France, is a vendor of hardware crypto-wallets. 
Their products are portable physical devices purposefully designed for \emph{offline storage} of cryptographic keys to crypto-assets.
Hardware wallets are widely advertised as the most secure option for managing digital assets~\cite{antonopoulos2014mastering}. 
However, upon device compromise---or in this case a long recovery phrase being disclosed---an attacker can get access to all crypto-assets secured by the device and transfer them irrevocably to accounts under his control.

In this market, Ledger competes with a handful of vendors selling similar security products. 
The vendor offers its customers a choice of two products, with basic or advanced functions, in the price range between \euro~80 and \euro~150. 
Given their technical features and purchase price, hardware crypto-wallets are often recommended for long-term storage, large crypto-asset portfolios, or users interested in additional security~\cite{AVBB2021-CHI,Nowroozi2023}. 

\begin{figure*}
	\input{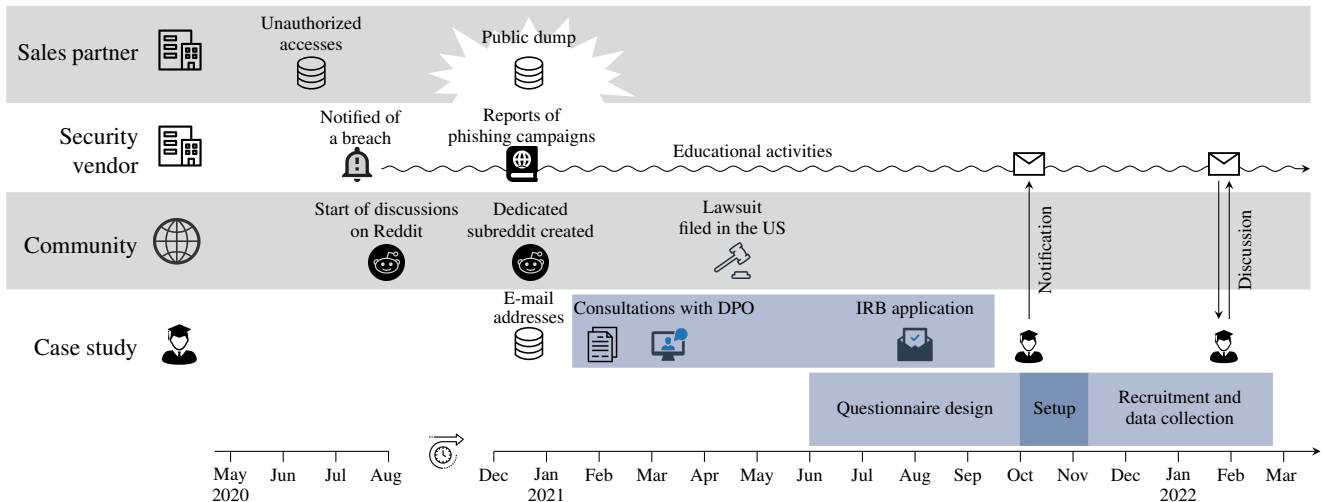}	
	\caption{Timeline of key events in our case study}\label{fig:timeline}
\end{figure*}

Figure~\ref{fig:timeline} shows a timeline of the key events of the breach and our study. 
Our reconstruction of the case is mainly sourced from information published by the companies involved~\cite{LedgerCISO2021,Shopify2020}, as independent and objective public information is rare.
Ledger partnered with Shopify, a popular e-commerce platform, to manage its online sales. 
In spring 2020, rogue employees at Shopify reportedly gained unauthorized access to customer records of up to 200 merchants~\cite{Shopify2020}, including Ledger~\cite{LedgerCISO2021}. 
In July 2020, a bug bounty hunter notified Ledger of a breach of their e-commerce and marketing database, involving a third-party API key. 
Ledger patched this problem and initiated internal investigations to estimate risks and protect its customers~\cite{LedgerCISO2021}. 
A subset of Ledger's marketing database, alleged to be the contents stolen earlier, was dumped in plain text on Raidforum on December 20, 2020. 
This public database contained personal information (first and last names, postal and e-mail addresses, phone numbers) of approximately 272\,000 customers~\cite{LedgerCISO2021}.

We got notified of this leak via social media and downloaded the dump from the Intelligence~X search engine and data archive\footnote{\url{https://intelx.io/}} on December 21, 2020. 
Given our purpose of use, we erased all personal data contained in the dump except for e-mail addresses of customers. 
We started the study by seeking legal advice form our university's data protection officer. 
For this, we outlined a document explaining our study, its methods as well as potential issues of concern. 
The design of a questionnaire started in June 2021, followed by several rounds of peer revision and pretests in October 2021. 
In parallel, we applied for ethical approval from the IRB in summer 2021. 
The recruitment of survey participants started on November 9, 2021 and continued until February 22, 2022. 
The data analysis and reporting of results were started and finalized in the summer and winter of 2022, respectively.

In the beginning of October 2021, before the start of the field phase, we notified Ledger about our study using the only e-mail address the security vendor provided on its website. The company's representatives established contact and discussed the study with us in January 2022, after they received reports about our running survey from the customers. 
We also shared a pre-print of the conditionally accepted paper in May 2023, allowing the company to prepare for potential customer reactions. 
Feedback received from the company at this stage, after cross-checking against independent sources, has helped us to clarify some facts.
We did not interact with Shopify.

The broader community of crypto-asset owners reacted to the breach on Reddit almost immediately after the vendor announced this fact on its website and in newsletter e-mails. 
When the dump got exposed to the public, the community created a subreddit forum dedicated to this incident. 
Also, a US-based law firm reportedly filed a class action complaint against the vendor in April 2021 \cite{SchneiderWallace2021}. 
The vendor has made efforts to support its customers by expanding their educational materials on security practices, shutting down phishing websites, and tracking scammers' transactions \cite{LedgerCISO2021}.

Note that we decided against using the community term `Ledger breach' in this paper. 
We are interested in studying the effects of this breach, but do not want to discredit any of the involved companies, which in this case reportedly fell victim to cybercriminals themselves.
\section{Related Work}\label{section:relwork}
Data breach research spans various disciplines, topics and methods. Historically, the most popular topics of study were breached organizations, their preventive and response measures~\cite{gwebu2018role,angst2017security}, costs~\cite{SWB2021-EDPL,romanosky2016examining},  stock market reactions~\cite{WB2021-SoK}, and workforce decisions~\cite{bana2022WEIShuman}. 
\citet{schlackl2022antecedents} present a systematic review of more than 120 articles published in 1990--2021 on a  range of the factors preceding and following a breach. One key takeaway of this work---underpinning our own view---is that this emerging research field lacks empirical studies measuring the effects of breaches on individuals, chiefly how much  real harm was incurred.

We conducted our own systematic search of potentially relevant articles in the ACM, IEEE, and information systems digital libraries, using a combination of key terms searched in titles or abstracts. For greater coverage, we also included a few academic venues which may solicit works related to data breaches. In total, we identified 92 papers of interest, out of which 20 can be found in the sample reviewed in \cite{schlackl2022antecedents}. 
Our own search reconfirms the ongoing evolution and growth of this research area. 
We summarize the most relevant findings here and refer the reader to \cite{schlackl2022antecedents} for a more detailed overview.

Topics of growing interest in the recent years include increased concerns~\cite{karunakaran2018data}, perceptions \cite{zou2018ve,mayer2021now,GAO2021102802,kude2017big,hassanzadeh2021user}, damages of different nature~\cite{bachura2022opm}, post-breach reactions and behaviors of victims~\cite{bhagavatula2020people,mayer2021now,turjeman2019data}. 
Methodologically, 
some studies survey general population panels to sample respondents with self-reported or verified breach experience~\cite{Ablon2016cons,mayer2021now}. 
Other works use hypothetical scenarios to frame potential effects of a breach and inquire individuals' behavioral reactions to it, although the respondents may never have experienced a breach event~\cite{karunakaran2018data}. 
A number of studies, like us, 
adopt a case-based approach to examining consumers' perspective on breaches. 
Selected examples include Equifax~\cite{zou2018ve}, the US Office of Personnel Management~\cite{bachura2022opm}, Ashley Madison~\cite{cross2019media}, Home Depot~\cite{SYED2019257}, and Canva~\cite{Kanva2021}. 
However, we could not find any study examining a breach affecting a security vendor. 
To the best of our knowledge, the present work is novel in this regard.

The results of prior studies indicate that individuals affected by a breach not always act on notifications or heed enclosed security advice~\cite{Ponemon2014,zou2018ve,zouCHI2020}. 
For example, only half of the population sample surveyed in \cite{Ablon2016cons} reported having changed passwords or personal identification numbers after being notified about an incident. 
Recent data on behavioral reactions to breaches look slightly better, with 88\% of the breach victims reporting to have changed their passwords in~\cite{mayer2021now}. 
However, the scope of this study spans a multitude of breaches of varying size and magnitude, which adds noise to the signal.  
Turning to explanatory factors, the general reluctance to make mitigation efforts can be explained partly by high perceived costs associated with protective actions~\cite{zou2018ve}, as well as low levels of concern for specific accounts or breached data items~\cite{Ablon2016cons,mayer2021now}. 
Against this backdrop, our work adds to the expanding body of knowledge by looking at a case involving victims with heightened security awareness~\cite{AVBB2021-CHI,lindqvist2021bitcoin} and concerns about their digital assets. 
Moreover, our study stands out for its analysis of more targeted and previously unseen threats. 
\section{Method}\label{section:method}
Surveying a sample of crypto-asset owners is known to be a challenging task in research \cite{AVBB2021-CHI}. 
The main obstacles include a globally disperse and heterogeneous population, the use of digital pseudonyms as user identifiers, privacy concerns, and the ensuing non-response bias. 
While commercial crowd-sourcing recruitment platforms extended their pre-screening filters with cryptocurrency-related fields and are being tested by scholars \cite{AVBB2021-CHI,mangipudi2022uncovering}, they still lack reliable and non-intrusive measures for the detection of fake self-reports of crypto-asset ownership. 
Against this backdrop, the leaked e-mail addresses offered us a unique research opportunity to recruit victims of the high-profile breach who are very likely to be genuine crypto-asset owners.

With ethical approval from our IRB and data protection officer, we sent out \emph{one-off}, \emph{non-personalized} e-mail invitations to a randomly chosen subset of the leaked addresses. 
Given a contentious nature of our recruitment approach and being guided by the principle of research transparency, we devote the next subsection to the discussion of legal and ethical aspects of our study. 
As we aim to demonstrate full transparency on our course of action, we specify technical and non-technical measures taken to minimize (additional) harm to the concerned parties. 
We then describe our recruitment and survey design processes in more detail (Sections~\ref{section:recruitment} and \ref{section:survey}), comment on the data quality (Section~ \ref{section:quality}), and describe the socio-demographics of our survey respondents (Section~\ref{section:profile}). 
\subsection{Legal and Ethical Aspects}\label{section:ethics}
The use of personal data---including nonconsensually acquired or leaked---has turned into a debated topic in the scientific community over the recent years. 
Many scholars, including us, concur that there is a scarcity of uniform guidelines, standards, or frameworks for a consistent assessment of legal and ethical risks of potentially problematic studies~\cite{Clark2019_ethics,Boustead2020ethics,Ienca2021ethical,Egelman2012_ethics}. 
This also holds true for security research, in which clear and enforceable codes of conduct are not developed yet ~\cite{Schrittwieser2013_ethics,Macnish2020}. 
The Menlo Report, a reference document intended to guide computer science research~\cite{2012-dittrich-mraf}, suggests to consult with a research ethics board when reusing existing data. 
The status quo assumes that scientists apply key principles for ethical research and identify benefits and potential harms of using data of illicit origin~\cite{2012-dittrich-mraf}, while ethics committees expose those studies to heightened scrutiny~\cite{Boustead2020ethics}.

Our study design needs to be examined in relation to both ethical and legal aspects. 
Under the EU General Data Protection Regulation (GDPR), e-mail addresses are personal data, the processing of which requires a legal basis. 
Consent is one possible legal basis. 
The nature of our study, however, rendered it infeasible to obtain prior consent from the victims for processing their e-mail addresses.
Therefore, we used this personal data item of the affected persons on the legal grounds of the public interest (Article 6, EU GDPR; with the additional safeguards detailed in Article 89, EU GDPR) and the freedom of scientific research (cf.~Article 13, EU Charter of Fundamental Rights). 
An argument in favor of using public interest as legal basis is the absence of alternative methods to contact a significant number of cryptocurrency owners, who are simultaneously customers of a security provider having suffered a data breach. 
This, in turn, is the prerequisite for our research that serves the public interest.

We use consent as a legal basis for all items collected after the initial contact, thereby adhering closely to the Menlo Report's guidelines \cite{2012-dittrich-mraf}. 
The first page of our survey asked for explicit consent to participation and personal data collection (see Appendix~\ref{app:questionnaire}). 
This page also included the purpose of this research, intended use, instructions, and contact details. 
The participants were informed of the anonymous nature of the data collection and an option to withdraw from the survey at any time during its completion. 
With this procedure, the results and anonymized comments of the participants reported in this paper are collected in compliance with the EU GDPR. 

Our review of potential ethical issues started with a search of guidelines, frameworks, and recommendations in the computer science literature \cite{Schrittwieser2013_ethics,Macnish2020,Thomas2017_ethics,2012-dittrich-mraf}. 
It became evident that most ethical discussions focus on the use case of analyzing data of illicit origin (e.\,g., user behavioral data) and inferring scientific knowledge directly from it. 
However, our intended purpose of use was to recruit survey participants by sending unsolicited invitations to the leaked e-mail addresses. 
This method of participant recruitment is not yet covered in the examined literature. 
We therefore followed the standard practice and did a risk--benefit analysis~\cite{2012-dittrich-mraf}. 

This study concerns the following groups of stakeholders: the victims themselves, the vendor, we as lead researchers, and the affiliated university. 
We report potential risks individually for each group, followed by mitigation strategies we put in place to minimize harm. 
With respect to the victims, the complete leaked dataset contained other sensitive personal information. 
Following the principle of data minimization, we permanently erased all data fields but the e-mail address. 
Furthermore, we accessed and stored the dataset in accordance with the EU guidelines for accessing confidential data for scientific purposes. 
We deleted the dataset and cleaned up our project mail account after the data collection had finished. 
All these security measures were meant to reduce the risk of another data leakage on our side.  
The victims could also have experienced additional distress or discomfort when receiving our one-off e-mail. 
Against this backdrop, we looked for conventional guidelines in the survey invitation design on how to reduce respondent burden \cite{einarsson2021reducing,kaufmann2020doing,brenner2020subject}, and elaborated the subject line and invitation text with succinct, but informative enough details about the purpose and nature of our study.
Victims who volunteered to respond to our survey have invested time and cognitive effort, without being compensated financially. 
We refrained from doing this to mitigate the risk of multiple participation,%
\footnote{The risk of multiple participation is significant in this target group. A blockchain analysis following the transaction that compensated 961 participants of~\cite{Krombholz2017} in bitcoins suggests that the total number of distinct entities is just above 600. One entity has submitted more than 50 responses.}
and the associated biases and response errors, in an anonymous online survey that intentionally did not track responses to invitations.

Our study could have harmed the vendor's reputation if it had found compromising results. 
To mitigate this potential harm, we informed the vendor of our research initiative before the start of participant recruitment (see Figure~\ref{fig:timeline}), and offered a communication channel. 
When presenting results, we refrain from exposing the vendor unnecessarily.
Note that we deliberately avoided a partnership with the vendor to ensure the independence and objectivity of this study. 
While the cooperation may have alleviated some ethical concerns, it would not have changed the legal basis of data processing as the breached organization had not obtained consent for the purpose of scientific research.  
More importantly, conducting an independent study allowed us to reduce the risk of uncontrollable response biases caused by potentially reduced consumer trust in the vendor. 
By doing so, we were also able to maintain the freedom of designing our own questionnaire and controlling the collected data.
	
As a final stakeholder, we or our university could be confronted with negative reactions or requests, including via e-mail communication or on social media platforms. 
We also identified the risk that our university domain could be blacklisted by providers abroad who do not share (or do not care about) our legal assessment of the e-mail distribution and thus (falsely) classify it as spam, or blindly follow takedown requests.
Throughout our study, we pursued the principle of being transparent and diligent when responding to any incoming or forwarded inquiry via our e-mail account, telephone, or other university staff members~\cite{2012-dittrich-mraf}. We kept track of a few social media posts discussing the study, however considered our direct intervention unnecessary.

We continue the ethical analysis with a list of potential benefits. 
First, the dumped data provided us with a unique opportunity to contact the victims and collect their valuable responses on the aftermath and response actions. 
In contrast to many other breaches, this incident predominantly concerned active or former cryptocurrency users. 
These persons are deemed to be a hard-to-reach community in behavioral security research~\cite{AVBB2021-CHI}. 
Therefore, neglecting this method of reaching out to potential research subjects would be a missed opportunity  in contributing empirical knowledge in this yet unexplored field.  
Second, we study the effects and behavioral changes caused by the breach exposure among the group of security-literate and aware individuals. 
Their responses will shed light on the scope and nature of harms even competent users cannot evade after a breach. 
By eliciting these empirical insights, we seek to add to the social good and advance user-centered breach research. 
Third, in the absence of financial compensation, this study will likely attract those participants who are interested in contributing to research and thus, willing to provide valuable responses. 

Overall, while we could not completely eliminate the risk of causing additional harm to the involved stakeholders, we implemented a series of adequate risk mitigation measures and safeguards for storing and processing e-mail addresses. 
We submitted the presented risk--benefit analysis as well as the designed questionnaire to the university's IRB and obtained a positive decision after one follow-up request. 
We kept in close contact with the IRB throughout the recruitment phase and informed it of the progress and received feedback.
\subsection{Participant Recruitment}\label{section:recruitment}
As already mentioned, we made use of the leaked data dump and extracted e-mail addresses to contact victims. 
Note that we refer to this reduced dataset of leaked addresses in the rest of this paper. 
The dataset was stored on an external secure USB flash drive kept in a safe vault. 
Access to it was granted to authorized members of the research group only. 
We connected the flash drive to a computer for querying the next batch of e-mail addresses in an offline mode only. 

Our recruitment strategy relied on \emph{non-personalized}, \emph{automated} e-mail invites which we distributed in batches to randomly selected addresses. 
Each batch included 2\,500 e-mails with a varying proportion of Gmail and non-Gmail accounts. 
We made this distinction in our e-mail sampling method after distributing the first batch and noticing that Gmail's built-in spam filters marked our invite as spam. 
We adjusted the header parameters in our mailing script (as we surmise that those most likely triggered anti-spam filters) and increased the number of Gmail accounts stepwise from 200 to 1250 in each following batch. 
We added four Gmail accounts under our control (in the beginning and at the end of each batch) in order to check for potential spam marking. 
In total, we initiated the distribution of 13 batches and sent out approximately 31\,632 e-mail invites 
(11.7\% of the leaked records).

Our approach to sample is a measure to reduce harm without severely compromising data quality.
Without knowledge of the response rate, we targeted a maximum sample size of 500 in the IRB request.
The actual collection was terminated when we had reached a number of complete responses that would guarantee at least 100 cases after data cleaning. 
This number can be justified with a statistical power analysis.\footnote{Specifically, 97 or more measurements are needed to have a confidence level of 95\% that the true value is within $\pm$10\%-pts.\ of the estimated value. This is relevant for point estimates of averages. Turning to rare events, with 90 or more samples we have 99\% probability to observe experiences at least once that apply to only 5\% of the population.\label{fn:power} By principle, power analysis is limited to the sampling error. The total measurement error is likely higher due to the coverage and non-response errors.} %

The e-mail invite, reproduced in Appendix~\ref{app:emailinvite}, contained a succinct description and purpose of the study. 
Following the GDPR requirements, it included information about the processing and storage of an e-mail address as well as the contact details of the principal investigator (PI) and data protection officer.
To reassure all contacted persons that the survey links are not personalized, we hosted a static webpage on our research group's website, which contained a short note explaining the purpose of this intermediary page as well as a link to the questionnaire generated by the host platform Qualtrics. 
Appendix~\ref{app:staticpage} reports the content of the static webpage.

\begin{figure*}[h]
	\centering
	\definecolor{maincolor}{HTML}{0000FF} 
	\input{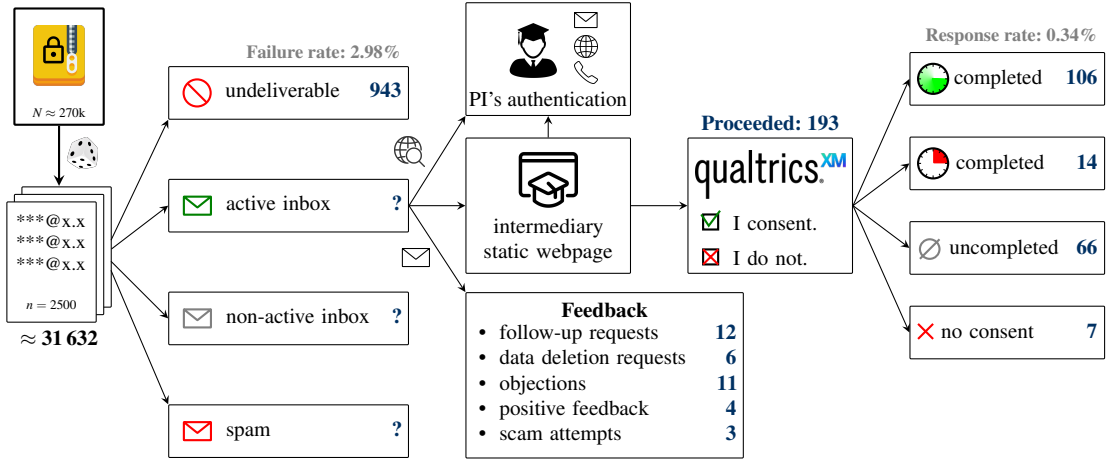}
	\vspace*{-8mm}
	\caption{Visualization of the participant recruitment workflow (numbers refer to cases in each branch)} \label{fig:recruitment}
\end{figure*}

Figure~\ref{fig:recruitment} presents the recruitment workflow and response statistics in greater detail.
In particular, we kept track of the  number of undeliverable messages which were bounced back to our university account due to abandoned, expired, full, or specially configured mailboxes. 
In total, we received 943 automatic non-delivery reports, equivalent to about 3\% of the total number of e-mails sent. 
Some of our e-mail invites may have ended up in spam or non-active inboxes, which we were not able to account for. 
According to the report provided by Qualtrics, 193 contacted persons proceeded to the introduction page of the questionnaire, out of which 186 consented and started the survey; 7 did not consent.

We received a number of follow-up messages requiring further information or an action on our side. 
A handful of the contacted individuals attempted to authenticate the lead researcher and used a range of creative strategies for that (e.\,g., via telephone calls or e-mails to a PI's business mailbox, e-mails to the research group's head or university staff members). 
Twelve persons inquired about the source from which the leaked addresses were collected, (neutrally) questioned the benevolent intentions or recruitment method of the study, or expressed their interest in the final results. 
There were 4 messages which explicitly provided positive feedback to our research initiative. 
By contrast, we received 6 and 11 requests exercising the right to erase and object to processing of the e-mail addresses, respectively, and 3 scam attempt messages. 
\subsection{Survey Instrument}\label{section:survey}
The process of designing a survey instrument was informed by multiple posts from affected victims on a dedicated Reddit discussion forum, which emerged after this incident. 
A non-systematic exploratory review of the posted messages revealed a set of attack vectors, response actions, and reported examples of harm, which were taken into account when phrasing  questions and answer options. 
While being guided by our primary research goals, we also borrowed some ideas and measurement scales from marketing, risk management, and other data breach studies~\cite{mayer2021now,masuch2020get,Ablon2016cons,ACSI2008}.

The questionnaire consisted of five parts.\ifquestionnaire{\protect\footnote{The complete questionnaire is available in Appendix~\ref{app:questionnaire}.}}{\protect\footnote{The complete questionnaire is available in an arXiv version of this paper.}}
\begin{description}
	\item[Part 1. Customer Experience] The survey started with an open-ended question eliciting {when} and {where} a participant had learned about the data breach. 
	The next block of questions addressed the purchase, use of, and satisfaction with a device, as well as the customers' expectations and attitudes toward the vendor and the product at the time before and after learning about the breach. 
We also asked participants whether they are aware of other hardware wallet vendors or even own their products. 
	\item[Part 2. Harm and Losses] This part focused on the breach's aftermath in terms of the harm caused to respondents, experienced attacks with monetary or other types of losses, and elevated concerns. 
	Since some effects may not necessarily have been caused by the data breach or in relation to it, we additionally asked respondents to evaluate their level of confidence in the causal relationship between this incident and the reported crime. 
	We also asked whether a respondent had fallen victim to another breach since January 2020. 
	\item[Part 3. Individual Response] Given our objective to explore individuals' security behaviors, we included a series of questions about precaution or protection actions the surveyed persons may have taken in response to this incident. 
	In order to capture a potential triggering effect of this event on a behavioral change, we explicitly asked in a number of questions whether a respondent had adopted a certain practice \emph{before} or \emph{after} the breach (or not at all).  
	This part also included questions on the discontinuance of using the leaked e-mail address or phone number, the purchased device, or any relationship with the vendor.
	\item[Part 4. Corporate Response] This part covered respondents' general expectations about corporate response strategies to a breach or theft of customer data. 
	We also attempted to discern customers' (dis)satisfaction with the actions taken by the vendor. 
	Since this part is related to the specifics of organizational responses and communication strategies, it is left out of scope of this paper.    
	\item[Part 5. Demographics] In this final part, we inquired about a respondent's age, gender, age, education, place of residence, and occupation. 
	In addition to the socio-demographic variables, we wished to enrich a descriptive profile of respondents with facts about their level of experience with crypto-assets measured in years and the approximate market value of crypto-assets owned. 
\end{description}

The online survey was hosted on the licensed Qualtrics survey platform. 
In order to protect participants' privacy, we disabled the registration of a respondent’s browser, operating system, IP address, or location information in the survey-specific settings. 
The survey could be accessed by anyone with a valid link. 
The estimated duration to complete the questionnaire was 25 minutes.
\subsection{Data Quality}\label{section:quality}
Several data quality measures were applied to ensure accurate and reliable responses. 
First, we reduced the dataset of completed responses ($N=120$) to 106 records which took 15 minutes or longer to complete. 
We set this time threshold to provide a reasonable balance between fast and slow respondents given the estimated completion time of 25 minutes.  
Then, we manually reviewed 7 data records which took between 14 and 15 minutes to complete and evaluated their quality. 
Out of these, we added 5 additional records based on our subjective judgment about their reliability. 
Given the relatively low number of responses, we also manually inspected open-ended text entry questions as well as response patterns for any presence of the non-differentiation, neutral answer selection, or primacy biases. 
As a result of this quality check, 7 records were excluded from the backbone set of 106 responses due to suspicious or non-meaningful answers.
After having applied these measures, the final dataset resulted into 104 responses, which are analyzed in this paper.

While clearly not representative for any meaningful population, we deem the collected data is of high reliability, since many respondents put extra effort into providing lengthy responses to open-ended questions. 
Respondents have spent an average of 34 minutes and a median of 26 minutes to fill out the survey. 
At the same time, we observe a high fraction of incomplete or explicit nondisclosure responses to the socio-demographic and profile questions. 
This confirms that crypto-asset owners tend to be privacy-conscious individuals who are not willing to disclose too many personal details about themselves.   
\subsection{Profile of Survey Respondents}\label{section:profile}
Table~\ref{table:demographics} summarizes the demographics of our sample. 
Most respondents are men in the age of 25 to 54 years, with a university degree and an employment status. 
Almost half of the sample have 3--4 years of experience with crypto-assets, while 40\% have been owners of crypto-assets for even longer. 
It comes at no surprise that almost half of the respondents prefer not to reveal the value of their investments. 
The ones who report (about one third of the sample) say that they hold crypto-assets worth between several thousands up to one million US dollars. (Using exchange rates at the time of completing the survey, not at the time of the breach.)
This confirms prior research on a broader sample, which revealed a highly significant relationship between owning more than \$10,000 in crypto-assets and the use of a hardware wallet~\cite[Table~9]{AVBB2021-CHI}. 

\begin{table}
	\caption{Demographics of the survey participants}\label{table:demographics}
	\scriptsize
	\begin{tabular}{@{}L{.31\columnwidth}@{}R{.07\columnwidth}@{}R{.09\columnwidth}@{\quad}L{.3\columnwidth}@{}R{.07\columnwidth}@{}R{.09\columnwidth}@{}}
		\toprule
		\multicolumn{2}{r@{}}{ \bf Abs.}&{ \bf \%}	&\multicolumn{2}{r@{}}{ \bf Abs.}&{ \bf \%}\\
		\midrule
		\multicolumn{3}{@{}L{.5\columnwidth}@{}}{ \bf Gender}&\multicolumn{1}{@{}L{.3\columnwidth}@{}}{ \bf Age}&&\\
		Men&88&84.6~&24 or younger&2&1.9~\\
		Women&2&1.9~&Between 25 and 34&20&19.2~\\
		Non-binary&2&1.9~&Between 35 and 44&24&23.1~\\
		I prefer not to answer.&12&11.5~&Between 45 and 54&26&25.0~\\
		&&&Between 55 and 64&10&9.6~\\
		&&&65 or older&7&6.7~\\
		&&&I prefer not to answer.&15&14.4~\\
		\midrule
		\multicolumn{3}{@{}L{.5\columnwidth}@{}}{ \bf Current occupation}&\multicolumn{1}{@{}L{.3\columnwidth}@{}}{ \bf Formal education}&&\\
		Student&0&0.0~&Less than high school&0&0.0~\\
		Skilled  manual worker&2&1.9~&High school incomplete&5&4.8~\\
		Employed in a service job&6&5.8~& High school graduate&11&10.6~\\
		Self-employed &14&13.5~& College or assoc. degree&14&13.5~\\
		Unemployed &2&1.9~& Bachelor's degree&25&24.0~\\
		Retired / sickness leave &9&8.7~& Master's degree&29&27.9~\\
		Employed professional &56&53.8~& Doctoral degree&6&5.8~\\
		Other &1&1.0~& Other professional degree&4&3.8~\\
		I prefer not to answer. &14&13.5~&I prefer not to answer.&10&9.6~\\
		\midrule
		\multicolumn{3}{@{}L{.5\columnwidth}@{}}{ \bf Experience with crypto-assets}&\multicolumn{3}{@{}L{.5\columnwidth}@{}}{ \bf Ownership of crypto-assets}\\
		Less than 1 year &0&0.0~&Less than \$1,000 &7&6.7~\\
		Between 1 and 2 years &12&11.5~&\$1,000 -- \$5,000 &5&4.8~\\
		Between 3 and 4 years &48&46.2~&\$5,000 -- \$10,000 &4&3.8~\\
		Between 5 and 6 years &18&17.3~&\$10,000 -- \$100,000 &18&17.3~\\
		More than 6 years &24&23.1~&\$100,000 -- \$1,000,000 &15&14.4~\\
		No answer&2&1.9~&	More than \$1,000,000 &4&3.8~\\
		&&&I prefer not to answer. &51&49.0~\\
		\bottomrule
	\end{tabular}
\end{table}

A total of 28 (26.9\%) participants report having been a victim of another data breach since January 2020 (43 no; and 33 not aware). The most popular reasons for purchasing a hardware wallet were 
\begin{enumerate*}[label=(\roman*)]
	\item to store crypto-assets for personal use (87.5\%),
	\item to give away as a gift (17.3\%),
	\item to review, experiment with and test security features (6.7\%), and
	\item to use for business purposes (5.8\%). 
\end{enumerate*}
More than half of the sample (56.7\%) report that they continue to use a purchased device, as opposed to those who never used it (11.5\%), or discontinued their usage before (3.8\%) or after (17.3\%) the breach. 
Out of those 12 respondents who never used the product, 6 purchased it in the hope of using it for personal needs, 3 as a giveaway or gift, and one for research. 

Besides Ledger, there are competing producers of hardware wallet devices (e.\,g., Trezor, Ellipal, or KeepKey). 
The majority of respondents (76\%) report not using similar products of other brands. 
The most common factors which respondents said influenced their decision to buy the Ledger product were
\begin{enumerate*}[label=(\roman*)]
	\item the vendor's reputation (75\%),
	\item product's technical security features (54.8\%),
	\item the number and type of supported crypto-assets and services (44.2\%), and
	\item recommendations from members of the community (39.4\%). 
\end{enumerate*}
These results suggest that Ledger's products are cherished and recognized by the global customer base for their security attributes and a broad and ever-growing range of supported crypto-assets.

\section{Empirical Results}\label{section:results}
 We structure the presentation of our main results along the research questions. 

\subsection{Experienced Harms and Attacks (RQ1)}
In general, victims of a data breach may anticipate or experience harms of different kinds. 
While acknowledging that beliefs about harm severity are often subjective and contingent on individual circumstances~\cite{mayer2021now}, we differentiate between serious damages and distress, such as account takeover attacks, physical threats, or money thefts; and less serious or negligible repercussions in form of scam, marketing, or nuisance e-mails, text messages, or phone calls. 
Figure~\ref{fig:harm} presents the frequency of experienced harms reported by the respondents. 
It should be emphasized that retrospective reports of the victims should be interpreted with caution due to potential recall bias and confounding factors, such as a respondent's involvement in another data breach or limited causal reasoning.

\begin{figure}[t]
	\centering
	\def\xmax{18}
	\def\ymax{100}
	\def\xmin{0.25}
	
	\begin{tikzpicture}[>=stealth,yscale=0.6*10/18, xscale=0.037,
	answer/.style={anchor=east, align=right, text width=3.8cm}]
	\foreach \x in {0, 10, 20, ...,90} {
		\draw[densely dotted,gray] (\x,0)--(\x,-14.75);
	}
	\begin{scope}[line width=10*10/18, altcolor!60, font = \scriptsize]
	\draw (0,-1)--(93*100/104,-1) node[right=-4pt, black] {93}
	(0,-2)--(82*100/104,-2) node[right=-4pt, black] {82}
	(0,-3)--(54*100/104,-3) node[right=-4pt, black] {54}
	(0,-4)--(33*100/104,-4) node[right=-4pt, black] {33}
	(0,-5)--(22*100/104,-5) node[right=-4pt, black] {22}
	(0,-6)--(19*100/104,-6) node[right=-4pt, black] {19}
	(0,-7)--(17*100/104,-7) node[right=-4pt, black] {17}
	(0,-8)--(6*100/104,-8) node[right=-4pt, black] {6}
	(0,-9)--(5*100/104,-9) node[right=-4pt, black] {5}
	(0,-10)--(4*100/104,-10) node[right=-4pt, black] {4}
	(0,-11)--(3*100/104,-11) node[right=-4pt, black] {3}
	(0,-12)--(2*100/104,-12) node[right=-4pt, black] {2}
	(0,-13)--(2*100/104,-13) node[right=-4pt, black] {2}
	(0,-14)--(2*100/104,-14) node[right=-4pt, black] {2}; 
	\end{scope}
	
	\foreach[count=\i from 1] \harm in {Scam \& phishing attacks, {Marketing e-mails \& text messages}, Password reset requests, Ransom threats, Scam physical letters, Attempt to takeover an account, Attempt to register a new account, New  registered account, \textbf{No harm at all}, Cyberbullying, SIM swap attacks,Successful account takover,Tampered wallet sent by post, Physical security threats}
	{
		\node[left,answer] at (0,-1*\i) {\scriptsize \harm \Aq};}
	
	\draw[<-] (100, -.25) node[above] {\scriptsize in \%} -- (0, -.25) -- (0, -14.75);
	\foreach \x in {0, 20, ...,80} {
		\draw (\x,-.25) --++(0,6pt) node[above=-2pt] {\scriptsize \x \%};}
	\foreach \x in {10, 30, ...,90} {\draw (\x,-.25) --++(0,6pt); }
	
	\end{tikzpicture}
	\caption{Reported harms and attacks (bars visualize percentages; numbers refer to absolute frequencies)}\label{fig:harm}
\end{figure}
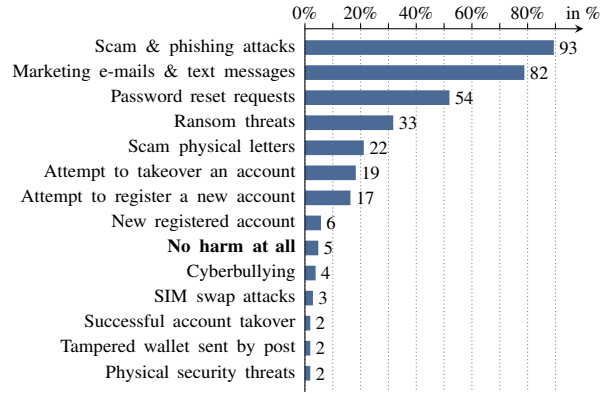

In line with prior research (e.\,g., \cite{mayer2021now}), the most prevalent forms of harm were scams, phishing attacks, marketing messages, and phone calls. 
Frequently, criminals were impersonating the vendor in an attempt to trick their target victims into sharing recovery phrases, updating devices, or installing malware on victim's computers. 
The victims were annoyed by targeted phishing e-mails and unsolicited marketing calls promoting crypto-asset investment opportunities. 
Almost half of the sample received password reset requests for accounts registered with a leaked e-mail. 
Criminal attempts to take over the control of online accounts were reported by 19 victims (18\%).
There are also 33 (32\%) reports of blackmailing and threats to disclose allegedly available personal data.

More serious forms of harm and advanced attack vectors seem to be rarer. 
Two instances of the successful account takeover were reported, while four persons received cyberbullying or physical threats. 
Two respondents reported receiving an authentically-looking parcel with a tampered crypto-wallet inside it, which was allegedly sent by the vendor with the security advice to transfer crypto-assets to the new device. 
While the numbers are too small for quantification attempts, we interpret this as existential evidence for resourceful attacks, requiring both technical expertise and upfront investments.

Out of the 104 responses, 10 reported financial losses which were incurred after July 2020. 
Figure~\ref{fig:finlosses} visualizes the magnitude of loss in a cryptocurrency, as reported by each participant. 
For the sake of comparability, we converted each number to its equivalent in US dollar by taking the average market price of a cryptocurrency in the second half of 2020. 
This approximation is dictated by the lack of exact timing information on the loss event. 
As visible in Figure~\ref{fig:finlosses}, most losses were attributed to scams and thefts (6 and 3 cases, respectively) and affected bitcoin and ether savings. 
However, according to the respondents' self-reports, only 4 events with losses ranging from one thousand up to several thousand US dollars are `likely' or `very likely' associated with the data breach. 
So, with some level of certainty, there were sporadic cases of successful cryptocurrency-related scams and thefts. 
Although the breach had no direct impact on the wallets and concerned sales information only, it provided criminals with channels to contact and take advantage of vulnerable victims. 

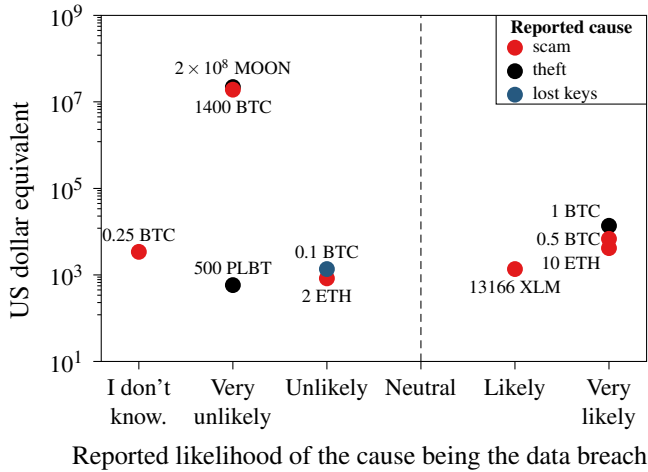
\begin{figure}[t]
	\begin{tikzpicture}[yscale=0.25,xscale=0.25]
	\draw (0,2.303)--(7.25*4,2.303)--(7.25*4,20.72)--(0,20.72)--cycle;
	\foreach \i/\l in {0.5*4/{I don't\Aq\\know.},1.75*4/{Very\Aq\\unlikely},3*4/{Unlikely},4.25*4/{Neutral},5.5*4/{Likely},6.75*4/{Very\Aq\\likely}} {
		\draw (\i,2.303)--++(0,-12pt) node[below, align=center, font = \small] {\l\Aq};
	}
	\foreach \y/\l in {2.303/$10^1$,6.908/{$10^3$},11.51/{$10^5$},16.12/{$10^7$}, 20.72/{$10^9$}} {
		\draw (0,\y)--++(-12pt,0) node [left, font = \small] {\l};
	}
	\foreach \y in {4.787,5.438,5.829,6.109,6.328,6.507,6.659,6.791,9.393,10.04,10.43,10.71,10.93,11.11,11.26,11.4,14,14.65,15.04,15.32,15.54,15.72,15.87,16,18.6,19.25,19.64,19.92,20.14,20.32,20.47,20.61}{
		\draw (0,\y)--++(-6pt,0);
	}
	\foreach \x/\y/\i/\c in {6.75*4/8.335/1/red_c, 0.5*4/8.136/2/red_c, 1.75*4/16.907/3/black,  5.5*4/7.220/4/red_c, 1.75*4/16.767/5/red_c, 6.75*4/9.523/6/black, 1.75*4/6.366/7/black, 3*4/6.726/8/red_c, 3*4/7.220/9/blue_c, 6.75*4/8.829/10/red_c}{
		\node[circle,fill=\c,inner sep=0pt,minimum size=6pt,] (n\i) at (\x,\y) {};
	}
	\begin{scope}[font=\scriptsize]
	\node[anchor=north east, inner sep = 3pt] at (n1) {10 ETH};
	\node[anchor=south, inner sep = 4pt] at (n2) {0.25 BTC};
	\node[anchor=south, inner sep = 4pt] at (n3) {${2\times10^8}$ MOON};
	\node[anchor=north, inner sep = 4pt] at (n4) {13166 XLM};
	\node[anchor=north, inner sep = 4pt] at (n5) {1400 BTC};
	\node[anchor=south east, inner sep = 3pt] at (n6) {1 BTC};
	\node[anchor=south, inner sep = 4pt] at (n7) {500 PLBT};
	\node[anchor=north, inner sep = 4pt] at (n8) {2 ETH};
	\node[anchor=south, inner sep = 4pt] at (n9) {0.1 BTC};
	\node[anchor=east, inner sep = 3pt] at (n10) { 0.5 BTC};
	\end{scope}
	\draw[densely dashed] (17,2.303)--(17,20.72);
	\node[rotate=90, yshift = 30 pt] at (0,20.72/2) {US dollar equivalent};
	\node[yshift=-20pt] at (13.75,0) { Reported likelihood of the cause being the data breach};
	
	\draw (7.25*4,16) --++(-8,0)--(21,20.72);
	
	\foreach \x/\y/\c/\l in {22/19/red_c/scam,22/17.75/black/theft,22/16.5/blue_c/{lost keys}}{\node[circle,fill=\c,inner sep=0pt,minimum size=6pt,label=right:\scriptsize\l\Aq] at (\x,\y) {};}
	
	\node at (25,20) {\scriptsize \bf Reported cause\Aq};
	\end{tikzpicture}
	\caption{Reported financial losses}\label{fig:finlosses}
\end{figure}

Besides asset security, physical safety risks (e.\,g., burglaries, assaults or kidnapping) could be of concern in the community of crypto-asset owners.  
Our data includes two reported instances of exposure to a physical threat. 
In response to another  question about individual's concerns, 30 respondents reported having moderate to extreme concerns about their physical safety. 
Quoting one respondent: \emph{``I used a temporary (exclusive to this vendor) e-mail and phone number for my order, but a postal address was genuine and thus presents the biggest risk for me and others working there.''} 

To sum up, the majority of the reported effects resembles those of typical data breaches and we find a few instances of threats which are characteristic to this case (e.\,g., fake physical devices, perceived safety hazard). 
At the same time, 5 respondents stated not having experienced any harm. 
\subsection{Response Actions (RQ2)}
As for the analysis of response actions, one should keep in mind that crypto-asset owners tend to be security-conscious, risk-aware, and proactive in taking protective actions~\cite{lindqvist2021bitcoin}.
This is, however, balanced by the growth of the global community of crypto-investors, which naturally becomes more heterogeneous~\cite{AVBB2021-CHI}. 
Looking at the socio-demographics of our sample (Table~\ref{table:demographics}), we observe that the majority of the surveyed victims have multiple years of experience with crypto-assets. 
It is safe to assume that they are familiar with basic security risks and protective measures relating to crypto-assets. 

To set the context, it is instructive to note that the large majority of victims in our sample has heard about the breach before suffering harm.
Almost half of the sample learned about it through the notification e-mail from the vendor or its newsletter, another 40\% from social media and online media;
only 7 individuals detected dubious activities first, and one person reportedly was hit unprepared and confronted with an empty wallet.
This means most response actions were post-breach, but still pro-active in relation to follow-up attacks.

Next we analyse from which sources of information the victims acquired help in relation to protective response actions. 
As shown in Figure~\ref{fig:response}, the survey participants largely took the vendor's security advice (45\%) or followed suggestions from online peer discussions (38\%).
This evidence hints at the importance of timely releases of advice on response strategies by breached organizations. 
It also suggests that opinion leaders in the crypto-asset community enjoy quite some trust.
 
	\begin{figure}[t]
		
		\centering
		\def\xmax{18}
		\def\ymax{100}
		\def\xmin{0.25}
		
		\begin{tikzpicture}[>=stealth,yscale=0.6*10/18, xscale=0.037,
		answer/.style={anchor=east, align=right, text width=4.75cm}]
		\foreach \x in {0, 10, 20, ...,60,70} {
			\draw[densely dotted,gray] (\x,-.25)--(\x,-16.75);
		}
		\begin{scope}[line width=10*10/18, altcolor!60, font = \scriptsize]
		\draw (0,-1)--(47*100/104,-1) node[right=-4pt, black] {47}
		(0,-2)--(40*100/104,-2) node[right=-4pt, black] {40}
		(0,-3)--(37*100/104,-3) node[right=-4pt, black] {37}
		(0,-4)--(33*100/104,-4) node[right=-4pt, black] {33}
		(0,-5)--(20*100/104,-5) node[right=-4pt, black] {20}
		(0,-6)--(17*100/104,-6) node[right=-4pt, black] {17}
		(0,-7)--(16*100/104,-7) node[right=-4pt, black] {16}
		(0,-8)--(13*100/104,-8) node[right=-4pt, black] {13}
		(0,-9)--(11*100/104,-9) node[right=-4pt, black] {11}
		(0,-10)--(9*100/104,-10) node[right=-4pt, black] {9}
		(0,-11)--(4*100/104,-11) node[right=-4pt, black] {4}
		(0,-12)--(3*100/104,-12) node[right=-4pt, black] {3}
		(0,-13)--(3*100/104,-13) node[right=-4pt, black] {3}
		(0,-14)--(2*100/104,-14) node[right=-4pt, black] {2}
		(0,-15)--(2*100/104,-15) node[right=-4pt, black] {2} 
		(0,-16)--(2*100/104,-16) node[right=-4pt, black] {2} ;
		\end{scope}
		
		\foreach[count=\i from 1] \harm in {Following the company's security advice, 
			Following social platform discussions,
			Resolving issues on my own, Browsing the company's website, Seeking for help online, Seeking for help from the community, No need for search, Sharing own experience, Seeking for help from my social circle, Seeking for help from the company, Seeking for help from security experts, I could not do anything., Managing spam \& phishing on my own, Contacting the police, Other, I don't know / recall.}
		{
			\node[left,answer] at (0,-1*\i) {\scriptsize \harm \Aq};}
		
		\draw[<-] (70, -.25) node[above, xshift=4pt] {\scriptsize in \%} -- (0, -.25) -- (0, -16.75);
		\foreach \x in {0, 20, ...,60} {
			\draw (\x,-.25) --++(0,6pt) node[above=-2pt] {\scriptsize \x \%};}
		\foreach \x in {10, 30, ...,50} {\draw (\x,-.25) --++(0,6pt); }
		
		\end{tikzpicture}

		\caption{Information search strategies used (bars visualize percentages; numbers refer to absolute frequencies)}\label{fig:response}
	\end{figure}
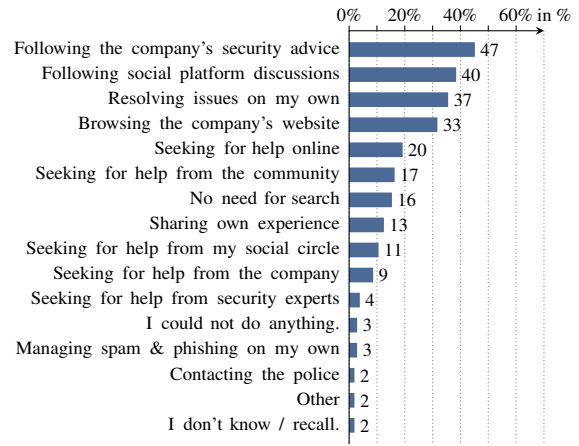

Figure~\ref{fig:protection} presents a list of the adopted protection measures ordered by the observed frequency. 
These represent actions the victims (reportedly) took in response to this breach. Some are of a general nature (e.\,g., using or extending the use of multi-factor authentication (MFA) or password managers), while others apply specifically to crypto-assets (e.\,g., changing passwords for crypto accounts or relocating backups of recovery seeds to a safer place). 
Overall, the most common practices were the increased use of MFA (52\%) or specialized authenticator apps (44\%), and the change of passwords for e-mail (47\%) and crypto-related accounts (41\%). 
In relation to the heightened safety concerns mentioned earlier, 18 respondents reported improving their home security. 

Interestingly, 22 respondents reported moving crypto-assets away from their crypto-wallets. 
This is a thought-provoking insight given that crypto-wallets are generally supposed to be secure and, more specifically, were not compromised by the breach. 
Unfortunately, our survey did not include a follow-up question on the nature and security level of the alternative storage options chosen by the victims who moved their funds. 
 
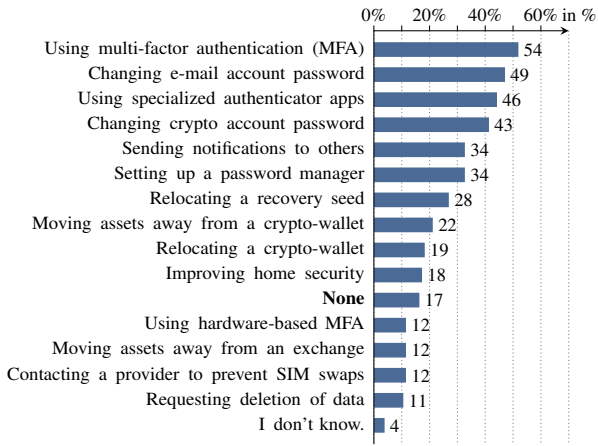
\begin{figure}[t]
	\centering
	\def\xmax{18}
	\def\ymax{100}
	\def\xmin{0.25}
	
	\begin{tikzpicture}[>=stealth,yscale=0.6*10/18, xscale=0.037,
	answer/.style={anchor=east, align=right, text width=4.75cm}]
	\foreach \x in {0, 10, 20, ...,60,70} {
		\draw[densely dotted,gray] (\x,-.25)--(\x,-16.75);
	}
	\begin{scope}[line width=10*10/18, altcolor!60, font = \scriptsize]
	\draw (0,-1)--(54*100/104,-1) node[right=-4pt, black] {54}
	(0,-2)--(49*100/104,-2) node[right=-4pt, black] {49}
	(0,-3)--(46*100/104,-3) node[right=-4pt, black] {46}
	(0,-4)--(43*100/104,-4) node[right=-4pt, black] {43}
	(0,-5)--(34*100/104,-5) node[right=-4pt, black] {34}
	(0,-6)--(34*100/104,-6) node[right=-4pt, black] {34}
	(0,-7)--(28*100/104,-7) node[right=-4pt, black] {28}
	(0,-8)--(22*100/104,-8) node[right=-4pt, black] {22}
	(0,-9)--(19*100/104,-9) node[right=-4pt, black] {19}
	(0,-10)--(18*100/104,-10) node[right=-4pt, black] {18}
	(0,-11)--(17*100/104,-11) node[right=-4pt, black] {17}
	(0,-12)--(12*100/104,-12) node[right=-4pt, black] {12}
	(0,-13)--(12*100/104,-13) node[right=-4pt, black] {12}
	(0,-14)--(12*100/104,-14) node[right=-4pt, black] {12}
	(0,-15)--(11*100/104,-15) node[right=-4pt, black] {11} 
	(0,-16)--(4*100/104,-16) node[right=-4pt, black] {4} ;
	\end{scope}

	\foreach[count=\i from 1] \harm in {Using multi-factor authentication (MFA), Changing e-mail account password,  Using specialized authenticator apps, Changing crypto account password, Sending notifications to others, Setting up a password manager, Relocating a recovery seed, Moving assets away from a crypto-wallet, Relocating a crypto-wallet, Improving home security, \textbf{None}, Using hardware-based MFA, Moving assets away from an exchange, Contacting a provider to prevent SIM swaps,  Requesting deletion of data, I don't know.}
	{
		\node[left,answer] at (0,-1*\i) {\scriptsize \harm \Aq};}
	
	\draw[<-] (70, -.25) node[above, xshift=4pt] {\scriptsize in \%} -- (0, -.25) -- (0, -16.75);
	\foreach \x in {0, 20, ...,60} {
		\draw (\x,-.25) --++(0,6pt) node[above=-2pt] {\scriptsize \x \%};}
	\foreach \x in {10, 30, ...,50} {\draw (\x,-.25) --++(0,6pt); }
	
	\end{tikzpicture}
	\caption{Adopted protection measures (bars visualize percentages; numbers refer to absolute frequencies)}\label{fig:protection}
\end{figure}

As a lesson learned from this case, the affected victims might have revised their security behaviors and adopted---or intended to adopt---measures which may reduce the probability of experiencing harm and its magnitude as an effect of a similar breach. 
For instance, one may allocate an e-mail address exclusive to crypto or financial operations, or use one-time addresses, postal boxes, or delivery forwarding services for online orders. 
To study these breach-specific implications, we showed respondents a list of 8 security practices and asked which they had practiced before or in response to the breach. 
If a person has not adopted a certain practice yet, we inquired the likelihood of doing so on a 5-point rating scale.
Figure~\ref{fig:precautions} summarizes the responses.

Probably because of its ease of implementation, the most-adopted action involves using dedicated e-mail addresses for any crypto-related business. 
In fact, managing multiple e-mail aliases and separating accounts for critical and general matters are what several respondents  emphasized retrospectively: \emph{``I am sad that I did not use a special e-mail address when purchasing the device.''} 
This is followed by the use of fake phone numbers or ``burner'' e-mail addresses for online orders. 
The use of fake names, decoy storage devices in case of a physical extortion of crypto-assets, or post office boxes for the delivery of online orders are less common, arguably due to practical limitations or inaccessibility. 
Finally, using postal forwarding services for the delivery of online orders or dedicated SIM cards for any business with crypto-assets were least adopted by the victims. 
When looking at the adoption intentions of the respondents who did not apply these measures yet, the overall patterns remain largely consistent with those of (self-reported) actual behaviors.  

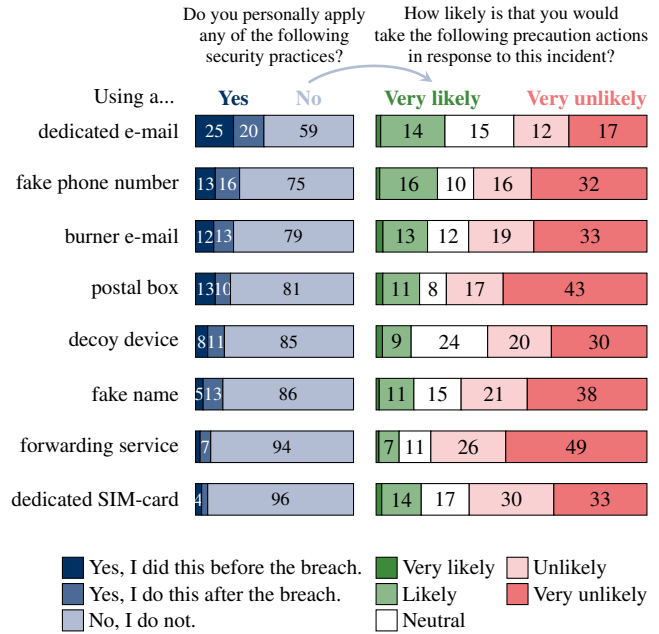
\begin{figure}
	\centering
	\definecolor{darkpurple}{HTML}{C51B7D}
	\definecolor{mediumpurple}{HTML}{DE77AE}
	\definecolor{mediumgreen}{HTML}{7FBC41}
	\definecolor{darkgreen}{HTML}{4D9221}

	\begin{tikzpicture}[yscale=0.7,xscale=0.6,
	choice/.style={},
	label/.style={},
	line_prec/.style={},
	line_prec_a/.style={fill=altcolor},
	line_prec_b/.style={fill=altcolor!60},
	line_prec_c/.style={fill=altcolor!20},
	label_prec/.style={color=white, font = \scriptsize},
	label_prec_a/.style={label_prec},
	label_prec_b/.style={label_prec},
	label_prec_c/.style={label_prec, color=black},
	line_int/.style={},
	line_int_a/.style={fill=green_c},
	line_int_b/.style={fill=green_c!60},
	line_int_c/.style={fill=white},
	line_int_d/.style={fill=red_c!20},
	line_int_e/.style={fill=red_c!60},
	label_int/.style={label},
	label_int_a/.style={label_int, black},
	label_int_b/.style={label_int, black},
	label_int_c/.style={label_int},
	label_int_d/.style={label_int, black},
	label_int_e/.style={label_int, black},
	callouts/.style={right,align=center,draw,fill=maincolor!20, drop shadow, rectangle callout},
	]
	\footnotesize
	\def\aax{0.2} 
	\def\ax{3.5} 
	\def\ay{0.6} 
	\def\abx{0.5} 
	\def\bx{6} 
	\begin{scope}[]
\draw[] (0,0) node[choice, anchor=east] (a){dedicated SIM-card};
\draw[line_prec_a] (a.east)++(\aax,\ay/2) coordinate(barA) rectangle node(yesb)[label_prec_a]{4} +(\ax*0.0385,-\ay) coordinate (b);
\draw[line_prec_b] (b) rectangle node(yesa)[label_prec_b]{} +(\ax*0.0385,\ay) coordinate (c);
\draw[line_prec_c] (c) rectangle node(no)[label_prec_c]{96} +(\ax*0.9231,-\ay) coordinate (d);
\draw[line_int_a] (d)++(\abx,0) coordinate(barB) rectangle node[label_int_a]{} +(\bx*0.0208,\ay) coordinate (e);
\draw[line_int_b] (e) rectangle node[label_int_b]{14} +(\bx*0.1458,-\ay) coordinate (f);
\draw[line_int_c] (f) rectangle node[label_int_c]{17} +(\bx*0.1771,\ay) coordinate (g);
\draw[line_int_d] (g) rectangle node[label_int_d]{30} +(\bx*0.3125,-\ay) coordinate (h);
\draw[line_int_e] (h) rectangle node[label_int_e]{33} +(\bx*0.3438,\ay) coordinate(i);

\draw[] (0,1) node[choice, anchor=east] (a){forwarding service};
\draw[line_prec_a] (a.east)++(\aax,\ay/2) coordinate(barA) rectangle node(yesb)[label_prec_a]{} +(\ax*0.0288,-\ay) coordinate (b);
\draw[line_prec_b] (b) rectangle node(yesa)[label_prec_b]{ 7} +(\ax*0.0673,\ay) coordinate (c);
\draw[line_prec_c] (c) rectangle node(no)[label_prec_c]{94} +(\ax*0.9038,-\ay) coordinate (d);
\draw[line_int_a] (d)++(\abx,0) coordinate(barB) rectangle node[label_int_a]{} +(\bx*0.0106,\ay) coordinate (e);
\draw[line_int_b] (e) rectangle node[label_int_b]{ 7} +(\bx*0.0745,-\ay) coordinate (f);
\draw[line_int_c] (f) rectangle node[label_int_c]{11} +(\bx*0.1170,\ay) coordinate (g);
\draw[line_int_d] (g) rectangle node[label_int_d]{26} +(\bx*0.2766,-\ay) coordinate (h);
\draw[line_int_e] (h) rectangle node[label_int_e]{49} +(\bx*0.5213,\ay) coordinate(i);

\draw[] (0,2) node[choice, anchor=east] (a){fake name};
\draw[line_prec_a] (a.east)++(\aax,\ay/2) coordinate(barA) rectangle node(yesb)[label_prec_a]{ 5} +(\ax*0.0481,-\ay) coordinate (b);
\draw[line_prec_b] (b) rectangle node(yesa)[label_prec_b]{13} +(\ax*0.1250,\ay) coordinate (c);
\draw[line_prec_c] (c) rectangle node(no)[label_prec_c]{86} +(\ax*0.8269,-\ay) coordinate (d);
\draw[line_int_a] (d)++(\abx,0) coordinate(barB) rectangle node[label_int_a]{} +(\bx*0.0116,\ay) coordinate (e);
\draw[line_int_b] (e) rectangle node[label_int_b]{11} +(\bx*0.1279,-\ay) coordinate (f);
\draw[line_int_c] (f) rectangle node[label_int_c]{15} +(\bx*0.1744,\ay) coordinate (g);
\draw[line_int_d] (g) rectangle node[label_int_d]{21} +(\bx*0.2442,-\ay) coordinate (h);
\draw[line_int_e] (h) rectangle node[label_int_e]{38} +(\bx*0.4419,\ay) coordinate(i);

\draw[] (0,3) node[choice, anchor=east] (a){decoy device};
\draw[line_prec_a] (a.east)++(\aax,\ay/2) coordinate(barA) rectangle node(yesb)[label_prec_a]{ 8} +(\ax*0.0769,-\ay) coordinate (b);
\draw[line_prec_b] (b) rectangle node(yesa)[label_prec_b]{11} +(\ax*0.1058,\ay) coordinate (c);
\draw[line_prec_c] (c) rectangle node(no)[label_prec_c]{85} +(\ax*0.8173,-\ay) coordinate (d);
\draw[line_int_a] (d)++(\abx,0) coordinate(barB) rectangle node[label_int_a]{} +(\bx*0.0235,\ay) coordinate (e);
\draw[line_int_b] (e) rectangle node[label_int_b]{ 9} +(\bx*0.1059,-\ay) coordinate (f);
\draw[line_int_c] (f) rectangle node[label_int_c]{24} +(\bx*0.2824,\ay) coordinate (g);
\draw[line_int_d] (g) rectangle node[label_int_d]{20} +(\bx*0.2353,-\ay) coordinate (h);
\draw[line_int_e] (h) rectangle node[label_int_e]{30} +(\bx*0.3529,\ay) coordinate(i);

\draw[] (0,4) node[choice, anchor=east] (a){postal box};
\draw[line_prec_a] (a.east)++(\aax,\ay/2) coordinate(barA) rectangle node(yesb)[label_prec_a]{13} +(\ax*0.1250,-\ay) coordinate (b);
\draw[line_prec_b] (b) rectangle node(yesa)[label_prec_b]{10} +(\ax*0.0962,\ay) coordinate (c);
\draw[line_prec_c] (c) rectangle node(no)[label_prec_c]{81} +(\ax*0.7788,-\ay) coordinate (d);
\draw[line_int_a] (d)++(\abx,0) coordinate(barB) rectangle node[label_int_a]{} +(\bx*0.0247,\ay) coordinate (e);
\draw[line_int_b] (e) rectangle node[label_int_b]{11} +(\bx*0.1358,-\ay) coordinate (f);
\draw[line_int_c] (f) rectangle node[label_int_c]{ 8} +(\bx*0.0988,\ay) coordinate (g);
\draw[line_int_d] (g) rectangle node[label_int_d]{17} +(\bx*0.2099,-\ay) coordinate (h);
\draw[line_int_e] (h) rectangle node[label_int_e]{43} +(\bx*0.5309,\ay) coordinate(i);

\draw[] (0,5) node[choice, anchor=east] (a){burner e-mail};
\draw[line_prec_a] (a.east)++(\aax,\ay/2) coordinate(barA) rectangle node(yesb)[label_prec_a]{12} +(\ax*0.1154,-\ay) coordinate (b);
\draw[line_prec_b] (b) rectangle node(yesa)[label_prec_b]{13} +(\ax*0.1250,\ay) coordinate (c);
\draw[line_prec_c] (c) rectangle node(no)[label_prec_c]{79} +(\ax*0.7596,-\ay) coordinate (d);
\draw[line_int_a] (d)++(\abx,0) coordinate(barB) rectangle node[label_int_a]{} +(\bx*0.0253,\ay) coordinate (e);
\draw[line_int_b] (e) rectangle node[label_int_b]{13} +(\bx*0.1646,-\ay) coordinate (f);
\draw[line_int_c] (f) rectangle node[label_int_c]{12} +(\bx*0.1519,\ay) coordinate (g);
\draw[line_int_d] (g) rectangle node[label_int_d]{19} +(\bx*0.2405,-\ay) coordinate (h);
\draw[line_int_e] (h) rectangle node[label_int_e]{33} +(\bx*0.4177,\ay) coordinate(i);

\draw[] (0,6) node[choice, anchor=east] (a){fake phone number};
\draw[line_prec_a] (a.east)++(\aax,\ay/2) coordinate(barA) rectangle node(yesb)[label_prec_a]{13} +(\ax*0.1250,-\ay) coordinate (b);
\draw[line_prec_b] (b) rectangle node(yesa)[label_prec_b]{16} +(\ax*0.1538,\ay) coordinate (c);
\draw[line_prec_c] (c) rectangle node(no)[label_prec_c]{75} +(\ax*0.7212,-\ay) coordinate (d);
\draw[line_int_a] (d)++(\abx,0) coordinate(barB) rectangle node[label_int_a]{} +(\bx*0.0133,\ay) coordinate (e);
\draw[line_int_b] (e) rectangle node[label_int_b]{16} +(\bx*0.2133,-\ay) coordinate (f);
\draw[line_int_c] (f) rectangle node[label_int_c]{10} +(\bx*0.1333,\ay) coordinate (g);
\draw[line_int_d] (g) rectangle node[label_int_d]{16} +(\bx*0.2133,-\ay) coordinate (h);
\draw[line_int_e] (h) rectangle node[label_int_e]{32} +(\bx*0.4267,\ay) coordinate(i);

\draw[] (0,7) node[choice, anchor=east] (a){dedicated e-mail};
\draw[line_prec_a] (a.east)++(\aax,\ay/2) coordinate(barA) rectangle node(yesb)[label_prec_a]{25} +(\ax*0.2404,-\ay) coordinate (b);
\draw[line_prec_b] (b) rectangle node(yesa)[label_prec_b]{20} +(\ax*0.1923,\ay) coordinate (c);
\draw[line_prec_c] (c) rectangle node(no)[label_prec_c]{59} +(\ax*0.5673,-\ay) coordinate (d);
\draw[line_int_a] (d)++(\abx,0) coordinate(barB) rectangle node[label_int_a]{} +(\bx*0.0169,\ay) coordinate (e);
\draw[line_int_b] (e) rectangle node[label_int_b]{14} +(\bx*0.2373,-\ay) coordinate (f);
\draw[line_int_c] (f) rectangle node[label_int_c]{15} +(\bx*0.2542,\ay) coordinate (g);
\draw[line_int_d] (g) rectangle node[label_int_d]{12} +(\bx*0.2034,-\ay) coordinate (h);
\draw[line_int_e] (h) rectangle node[label_int_e]{17} +(\bx*0.2881,\ay) coordinate(i);
	\end{scope}
	
	\draw (barA)+(\ax/2,1.5) node[align=center, font = \scriptsize] {Do you personally apply\\ any of the following\\ security practices?};
	\draw (barB)+(\bx/2,\ay+1.5) node(question2)[align=center, font = \scriptsize] {How likely is that you would\\ take the following precaution actions\\ in response to this incident?};

	\draw (b)+(-2.2,0.33+\ay)  node {Using a...};
	\draw (b)+(0,0.33+\ay) node[altcolor]{\bf Yes};
	\draw (no)+(0,0.33+\ay/2) node[altcolor!20,inner sep=0,outer sep=3pt] (nolabel) {\bf No};
	
	\draw (i)+(0,0.3) node[left, red_c!60,inner sep=0,outer sep=0]{\bf Very unlikely};
	\draw (d)+(\abx,\ay+0.3)node [right,green_c,inner sep=0,outer sep=3pt] (vlikelylabel) {\bf Very likely};
	
	\draw[->, >=stealth, altcolor!20, line width = 1pt] (nolabel.north) to [bend left,looseness=.75] (vlikelylabel.north);
	
	\draw[line_prec_a] (-2.75,-1.5) rectangle ++(0.4*7/6,0.4);
	\draw[line_prec_b] (-2.75,-2) rectangle ++(0.4*7/6,0.4);
	\draw[line_prec_c] (-2.75,-2.5) rectangle ++(0.4*7/6,0.4);

	\begin{scope}[choice, anchor=west, inner sep = 0pt, yshift=-1pt]
			\node at (-2.18,-1.3) {Yes, I did this before the breach.\Aq};
			\node at (-2.18,-1.8) {Yes, I do this after the breach.\Aq};
			\node at (-2.18,-2.3) {No, I do not.\Aq};
			\node (unl) at (4.77,-1.3) {Very likely \Aq};
			\node at (4.77,-1.8) {Likely \Aq};
			\node at (4.77,-2.3) {Neutral \Aq};
			\node at (7.67,-1.3) {Unlikely \Aq};
			\node at (7.67,-1.8) {Very unlikely \Aq};
	\end{scope}
	
	\draw[line_int_a] (4.2,-1.5) rectangle ++(0.4*7/6,0.4);
	\draw[line_int_b] (4.2,-2) rectangle ++(0.4*7/6,0.4);
	\draw[line_int_c] (4.2,-2.5) rectangle ++(0.4*7/6,0.4);
	
	\draw[line_int_d] (7.1,-1.5) rectangle ++(0.4*7/6,0.4);
	\draw[line_int_e] (7.1,-2) rectangle ++(0.4*7/6,0.4);
	\end{tikzpicture}
	\caption{Security practices and adoption intentions (numbers refer to absolute frequencies)}\label{fig:precautions}
\end{figure}

Seeking for legal protection and compensation is another possible response action to a breach.  
Two persons in our sample reported being part of a class-action lawsuit against the security vendor.\footnote{We are not aware of class actions against the sales partner and did not offer this answer option. Future survey instruments could improve by considering all involved parties as potential defendants.}
Many open-ended comments on this matter were along the lines of  \emph{``too much effort with little perspective of success''}, \emph{``too expensive''}, \emph{``time and money needed would not hold up against potential benefits''}, or \emph{``harm is already done.''} 
So, victims seem to be rather skeptical about pursuing litigation against breached entities. 
Specifically to our case, this skepticism could have been caused by difficulties of proving asset losses or suffering from cognizable damage as a result of the breach.

\subsection{Consumer Attitudes (RQ3)}
Our review of Reddit messages has shown a salient divergence of the post-breach attitudes of victims towards the security vendor as compared to its products. 
There were cases of negative reactions to the incident, blaming the vendor for failing to protect customer data. 
At the same time, some consumers publicly admitted that while they may no longer trust the vendor, the purchased product continues to meet their needs and original expectations: \emph{``[$\ldots$] I have no issue with the devices and consider them to be solid for the purpose I use them.''\footnote{\url{https://www.reddit.com/r/ledgerwalletleak/comments/kpzm4x/is\_anyone\_still\_using\_ledger\_had\_a\_close\_call/}}}

A potential divergence in attitudes towards a company's societal obligations and product expertise has been a topic of interest in the marketing literature for many years~\cite{brown1997company}. 
Specifically, individuals' corporate image associations are empirically found to influence---sometimes counter-intuitively---product evaluations~\cite{gurhan2004corporate,brown1997company}. 
Being inspired by this firm-versus-product level of abstraction, we asked respondents to rate their attitudes towards the vendor and, separately, the security device using 5-point rating scales (1 -- very negative, 5 -- very positive). 
In order to track any attitudinal changes triggered by the breach, the question wording explicitly invited the participants to recollect their attitudes at the time before and after the breach.

Figure~\ref{fig:sankey-attitudes} summarizes the observed patterns in two Sankey diagrams, showing the apparent changes in individuals' attitudes. 
Not very surprisingly, almost half of our sample reports \emph{negative} or \emph{very negative} post-breach attitudes towards the vendor. 
Likewise, only 6\% out of 65\% of those who recollect to have a very positive attitude prior to the incident remained loyal in their ex-post evaluation.  
On the other hand, the breach had a moderate effect on the consumers' attitudes towards the product: there are no drastic changes between the different levels of the attitude scale. 
Statistically, a Wilcoxon rank sum test confirms that the differences in attitudes towards the vendor are highly significant ($p<0.001$), whereas the attitudes towards the product do not differ significantly ($p=0.095$).   

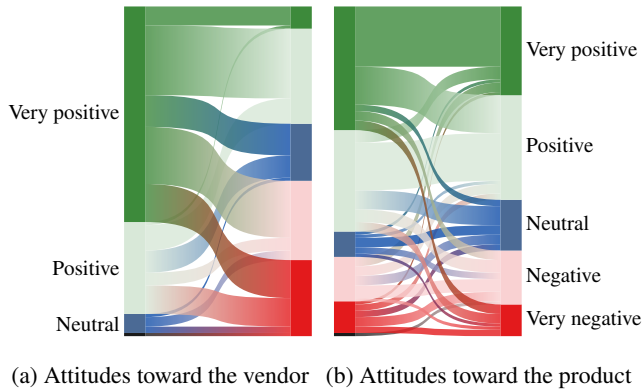
\begin{figure}[t]
	\centering
	\begin{subfigure}[t]{0.5\columnwidth}
		\centering
		\begin{tikzpicture}[y=1.2 pt,x=55pt]
		\begin{scope}[opacity=.8]
		\path [shade,left color=IdkColor, right color=VeryNegativeColor] (0,1) .. controls (0.75,1) and (0.25,1) .. (1,1)--(1,0) .. controls (0.25,0) and (0.75,0) .. (0,0)--cycle;
\path [shade,left color=NeutralColor, right color=VeryNegativeColor] (0,3) .. controls (0.75,3) and (0.25,3) .. (1,3)--(1,1) .. controls (0.25,1) and (0.75,1) .. (0,1)--cycle;
\path [shade,left color=NeutralColor, right color=NegativeColor] (0,6) .. controls (0.75,6) and (0.25,27) .. (1,27)--(1,24) .. controls (0.25,24) and (0.75,3) .. (0,3)--cycle;
\path [shade,left color=NeutralColor, right color=NeutralColor] (0,7) .. controls (0.75,7) and (0.25,50) .. (1,50)--(1,49) .. controls (0.25,49) and (0.75,6) .. (0,6)--cycle;
\path [shade,left color=PositiveColor, right color=VeryNegativeColor] (0,16) .. controls (0.75,16) and (0.25,12) .. (1,12)--(1,3) .. controls (0.25,3) and (0.75,7) .. (0,7)--cycle;
\path [shade,left color=PositiveColor, right color=NegativeColor] (0,20) .. controls (0.75,20) and (0.25,31) .. (1,31)--(1,27) .. controls (0.25,27) and (0.75,16) .. (0,16)--cycle;
\path [shade,left color=PositiveColor, right color=NeutralColor] (0,27) .. controls (0.75,27) and (0.25,57) .. (1,57)--(1,50) .. controls (0.25,50) and (0.75,20) .. (0,20)--cycle;
\path [shade,left color=PositiveColor, right color=PositiveColor] (0,35) .. controls (0.75,35) and (0.25,75) .. (1,75)--(1,67) .. controls (0.25,67) and (0.75,27) .. (0,27)--cycle;
\path [shade,left color=PositiveColor, right color=VeryPositiveColor] (0,36) .. controls (0.75,36) and (0.25,98) .. (1,98)--(1,97) .. controls (0.25,97) and (0.75,35) .. (0,35)--cycle;
\path [shade,left color=VeryPositiveColor, right color=VeryNegativeColor] (0,48) .. controls (0.75,48) and (0.25,24) .. (1,24)--(1,12) .. controls (0.25,12) and (0.75,36) .. (0,36)--cycle;
\path [shade,left color=VeryPositiveColor, right color=NegativeColor] (0,66) .. controls (0.75,66) and (0.25,49) .. (1,49)--(1,31) .. controls (0.25,31) and (0.75,48) .. (0,48)--cycle;
\path [shade,left color=VeryPositiveColor, right color=NeutralColor] (0,76) .. controls (0.75,76) and (0.25,67) .. (1,67)--(1,57) .. controls (0.25,57) and (0.75,66) .. (0,66)--cycle;
\path [shade,left color=VeryPositiveColor, right color=PositiveColor] (0,98) .. controls (0.75,98) and (0.25,97) .. (1,97)--(1,75) .. controls (0.25,75) and (0.75,76) .. (0,76)--cycle;
\path [shade,left color=VeryPositiveColor, right color=VeryPositiveColor] (0,104) .. controls (0.75,104) and (0.25,104) .. (1,104)--(1,98) .. controls (0.25,98) and (0.75,98) .. (0,98)--cycle;
		\end{scope}
		
		
		\begin{scope}[line width=8pt]
		\footnotesize
		\coordinate (Q) at (-4pt,0);
		\draw [IdkColor] (Q)-- node [left,black] {} ++(0,1) coordinate (Q);
\draw [NeutralColor] (Q)-- node [left,black, xshift=1pt] {Neutral} ++(0,6) coordinate (Q);
\draw [PositiveColor] (Q)-- node [left,black, xshift=1pt] {Positive} ++(0,29) coordinate (Q);
\draw [VeryPositiveColor] (Q)-- node [left,black, align=right, xshift=1pt] {Very positive} ++(0,68) coordinate (Q);
		\end{scope}
		
		
		\begin{scope}[line width=8pt]
		\footnotesize
		\draw (1,0)++(4pt,0) coordinate (Q);
\draw [VeryNegativeColor] (Q)-- node [right,black,align=left, xshift=-1pt] {} ++(0,24) coordinate (Q);
\draw [NegativeColor] (Q)-- node [right,black, xshift=-1pt] {} ++(0,25) coordinate (Q);
\draw [NeutralColor] (Q)-- node [right,black, xshift=-1pt] {} ++(0,18) coordinate (Q);
\draw [PositiveColor] (Q)-- node [right,black, xshift=-1pt] {} ++(0,30) coordinate (Q);
\draw [VeryPositiveColor] (Q)-- node [right,black,pos=0.5, xshift=-1pt] { } ++(0,7) coordinate (Q);	
		\end{scope}
		\end{tikzpicture}
		\subcaption{Attitudes toward the vendor}\label{fig:sankey-firm-attit}
	\end{subfigure}%
	\begin{subfigure}[t]{0.5\columnwidth}
		\centering
		\begin{tikzpicture}[y=1.2pt,x=55pt]
		\begin{scope}[opacity=.8]
		\path [shade,left color=IdkColor, right color=NegativeColor] (0,1) .. controls (0.75,1) and (0.25,11) .. (1,11)--(1,10) .. controls (0.25,10) and (0.75,0) .. (0,0)--cycle;
\path [shade,left color=VeryNegativeColor, right color=VeryNegativeColor] (0,3) .. controls (0.75,3) and (0.25,2) .. (1,2)--(1,0) .. controls (0.25,0) and (0.75,1) .. (0,1)--cycle;
\path [shade,left color=VeryNegativeColor, right color=NegativeColor] (0,6) .. controls (0.75,6) and (0.25,14) .. (1,14)--(1,11) .. controls (0.25,11) and (0.75,3) .. (0,3)--cycle;
\path [shade,left color=VeryNegativeColor, right color=NeutralColor] (0,8) .. controls (0.75,8) and (0.25,29) .. (1,29)--(1,27) .. controls (0.25,27) and (0.75,6) .. (0,6)--cycle;
\path [shade,left color=VeryNegativeColor, right color=PositiveColor] (0,10) .. controls (0.75,10) and (0.25,45) .. (1,45)--(1,43) .. controls (0.25,43) and (0.75,8) .. (0,8)--cycle;
\path [shade,left color=VeryNegativeColor, right color=VeryPositiveColor] (0,11) .. controls (0.75,11) and (0.25,77) .. (1,77)--(1,76) .. controls (0.25,76) and (0.75,10) .. (0,10)--cycle;
\path [shade,left color=NegativeColor, right color=VeryNegativeColor] (0,12) .. controls (0.75,12) and (0.25,3) .. (1,3)--(1,2) .. controls (0.25,2) and (0.75,11) .. (0,11)--cycle;
\path [shade,left color=NegativeColor, right color=NegativeColor] (0,16) .. controls (0.75,16) and (0.25,18) .. (1,18)--(1,14) .. controls (0.25,14) and (0.75,12) .. (0,12)--cycle;
\path [shade,left color=NegativeColor, right color=NeutralColor] (0,19) .. controls (0.75,19) and (0.25,32) .. (1,32)--(1,29) .. controls (0.25,29) and (0.75,16) .. (0,16)--cycle;
\path [shade,left color=NegativeColor, right color=PositiveColor] (0,22) .. controls (0.75,22) and (0.25,48) .. (1,48)--(1,45) .. controls (0.25,45) and (0.75,19) .. (0,19)--cycle;
\path [shade,left color=NegativeColor, right color=VeryPositiveColor] (0,25) .. controls (0.75,25) and (0.25,80) .. (1,80)--(1,77) .. controls (0.25,77) and (0.75,22) .. (0,22)--cycle;
\path [shade,left color=NeutralColor, right color=VeryNegativeColor] (0,26) .. controls (0.75,26) and (0.25,4) .. (1,4)--(1,3) .. controls (0.25,3) and (0.75,25) .. (0,25)--cycle;
\path [shade,left color=NeutralColor, right color=NegativeColor] (0,28) .. controls (0.75,28) and (0.25,20) .. (1,20)--(1,18) .. controls (0.25,18) and (0.75,26) .. (0,26)--cycle;
\path [shade,left color=NeutralColor, right color=NeutralColor] (0,31) .. controls (0.75,31) and (0.25,35) .. (1,35)--(1,32) .. controls (0.25,32) and (0.75,28) .. (0,28)--cycle;
\path [shade,left color=NeutralColor, right color=PositiveColor] (0,32) .. controls (0.75,32) and (0.25,49) .. (1,49)--(1,48) .. controls (0.25,48) and (0.75,31) .. (0,31)--cycle;
\path [shade,left color=NeutralColor, right color=VeryPositiveColor] (0,33) .. controls (0.75,33) and (0.25,81) .. (1,81)--(1,80) .. controls (0.25,80) and (0.75,32) .. (0,32)--cycle;
\path [shade,left color=PositiveColor, right color=VeryNegativeColor] (0,36) .. controls (0.75,36) and (0.25,7) .. (1,7)--(1,4) .. controls (0.25,4) and (0.75,33) .. (0,33)--cycle;
\path [shade,left color=PositiveColor, right color=NegativeColor] (0,40) .. controls (0.75,40) and (0.25,24) .. (1,24)--(1,20) .. controls (0.25,20) and (0.75,36) .. (0,36)--cycle;
\path [shade,left color=PositiveColor, right color=NeutralColor] (0,46) .. controls (0.75,46) and (0.25,41) .. (1,41)--(1,35) .. controls (0.25,35) and (0.75,40) .. (0,40)--cycle;
\path [shade,left color=PositiveColor, right color=PositiveColor] (0,61) .. controls (0.75,61) and (0.25,64) .. (1,64)--(1,49) .. controls (0.25,49) and (0.75,46) .. (0,46)--cycle;
\path [shade,left color=PositiveColor, right color=VeryPositiveColor] (0,65) .. controls (0.75,65) and (0.25,85) .. (1,85)--(1,81) .. controls (0.25,81) and (0.75,61) .. (0,61)--cycle;
\path [shade,left color=VeryPositiveColor, right color=VeryNegativeColor] (0,68) .. controls (0.75,68) and (0.25,10) .. (1,10)--(1,7) .. controls (0.25,7) and (0.75,65) .. (0,65)--cycle;
\path [shade,left color=VeryPositiveColor, right color=NegativeColor] (0,71) .. controls (0.75,71) and (0.25,27) .. (1,27)--(1,24) .. controls (0.25,24) and (0.75,68) .. (0,68)--cycle;
\path [shade,left color=VeryPositiveColor, right color=NeutralColor] (0,73) .. controls (0.75,73) and (0.25,43) .. (1,43)--(1,41) .. controls (0.25,41) and (0.75,71) .. (0,71)--cycle;
\path [shade,left color=VeryPositiveColor, right color=PositiveColor] (0,85) .. controls (0.75,85) and (0.25,76) .. (1,76)--(1,64) .. controls (0.25,64) and (0.75,73) .. (0,73)--cycle;
\path [shade,left color=VeryPositiveColor, right color=VeryPositiveColor] (0,104) .. controls (0.75,104) and (0.25,104) .. (1,104)--(1,85) .. controls (0.25,85) and (0.75,85) .. (0,85)--cycle;
		\end{scope}
		
		
		\begin{scope}[line width=8pt]
		\footnotesize
		\coordinate (Q) at (-4pt,0);
		\draw [IdkColor] (Q)-- node [left,black] {} ++(0,1) coordinate (Q);
\draw [VeryNegativeColor] (Q)-- node [left,black,pos=0.5, xshift=1pt] {} ++(0,10) coordinate (Q);
\draw [NegativeColor] (Q)-- node [left,black, xshift=1pt] {} ++(0,14) coordinate (Q);
\draw [NeutralColor] (Q)-- node [left,black, xshift=1pt] {} ++(0,8) coordinate (Q);
\draw [PositiveColor] (Q)-- node [left,black, xshift=1pt] {} ++(0,32) coordinate (Q);
\draw [VeryPositiveColor] (Q)-- node [left,black,align=center, xshift=1pt] {} ++(0,39) coordinate (Q);
		\end{scope}
		
		
		\begin{scope}[line width=8pt]
		\footnotesize
		\draw (1,0)++(4pt,0) coordinate (Q);
\draw [VeryNegativeColor] (Q)-- node [right,black,pos=0.5, xshift=-1pt] {Very negative\Aq} ++(0,10) coordinate (Q);
\draw [NegativeColor] (Q)-- node [right,black, xshift=-1pt] {Negative\Aq} ++(0,17) coordinate (Q);
\draw [NeutralColor] (Q)-- node [right,black, xshift=-1pt] {Neutral\Aq} ++(0,16) coordinate (Q);
\draw [PositiveColor] (Q)-- node [right,black, xshift=-1pt] {Positive\Aq} ++(0,33) coordinate (Q);
\draw [VeryPositiveColor] (Q)-- node [right,black, xshift=-1pt] {Very positive\Aq} ++(0,28) coordinate (Q);	
		\end{scope}
		\end{tikzpicture}
		\subcaption{Attitudes toward the product}\label{fig:sankey-prod-attit}
	\end{subfigure}%
	\caption{Consumer attitudes \emph{before} 
	and \emph{after} the breach}\label{fig:sankey-attitudes}
\end{figure}
\section{Discussion}\label{section:discussion}
This case study sheds light on a high-profile data breach which exposed a population of affluent crypto-asset owners as potentially attractive targets for opportunistic profit-oriented crimes. 
In line with prior work~\cite{karunakaran2018data,zou2018ve,mayer2021now}, our findings show that scam, phishing and attempted identity thefts characterize the aftermaths of breaches. 
This specific case reveals a physical attack vector previously unseen in data breach research (to the best of our knowledge). 
It could explain our finding of elevated safety concerns among the victims whose residence addresses were leaked. 
As some respondents pointed out: changing a place of living is not as easy as switching to a new e-mail address. 

In Section~\ref{section:implications}, we discuss the implications of our study and its key findings for the cybersecurity industry and related policy initiatives. 
Next, we share our lessons learned regarding the methodological challenges in data breach research (Section~\ref{section:implresearch}). 
Section~ \ref{section:limitations} addresses potential limitations.
\subsection{Implications for Industry and Policy}\label{section:implications}
The victims in our sample were exposed to a range of crimes exploiting almost all leaked data items. 
Most reported attacks were targeted, domain-specific, and plausibly related to the breach event, even though our method cannot establish causality. 
This led us to the insight that breaches affecting security vendors (and by extension security service providers) deserve special attention in data breach research.  
We call these breaches `high-profile' and note their potentially far-reaching effects because they enable targeted follow-up crimes. 
In this sense, the identification of affluent crypto-asset holders as users of a Ledger wallet is qualitatively similar to the exposure of website user accounts in the recent LastPass case.

Beyond doubt, breaches will continue to plague the world.  
Against this background, it is essential like never before that the cybersecurity industry serves as a role model of effective information security and crisis management. 
After all, companies whose mission is to provide excellence in security products and services should stand by their own promises. 
In the age of digital transformation, these expectations add a new form of corporate social responsibility, which, as this case has shown, extends beyond the boundaries of the security vendor. 
In particular security companies should exercise extreme caution when choosing business partners and avoid sales channels not designed for dealing with the risk of a high-profile breach. 
Several of our findings reveal a potential disconnect between victims' concerns and material harm suffered.
This underscores the importance of effective corporate crisis communication and the management of reputation threats, in particular on social media platforms~\cite{syed2019enterprise}.

When looking at the anatomy of the case at large, we conjecture that more harm was prevented through perceptions and mitigating actions of victims and by the vendor. 
Owners of crypto-wallets are reputedly the most mistrustful among the already security-concerned crowd of crypto-asset owners. 
This trait, which we also observed during recruitment, further connects to a found ability of the surveyed participants to evaluate the security properties of products independent of a vendor's reputation. 
While this may be specific to the target group of our study, it may call for a critical (and empirical) reassessment of a commonly accepted fact in the security economics literature, namely that consumers cannot evaluate security properties when making purchase decisions~\cite{anderson2001information}.

Stricter vendor liability is often postulated as a cure for market failures and as a means to incentivize security practices and investments~\cite{anderson1994liability,laszka2014survey}. 
For example, the 2023 US National Cybersecurity Strategy sets out to ``reshape laws that govern liability for data losses and harm caused by cybersecurity errors''~\cite[p.~19]{whitehouse2023}.
However, the liability channel is prone to fail if victims keep skeptical about its success and do not support it wholeheartedly, as our results have demonstrated (albeit under the current lax liability regime).  
This calls for studying the subjective causes of perceived skepticism towards litigation and a potential reform of cybersecurity law~\cite{kilovaty2021psychological}. 

Prior work~\cite{mayer2021now} puts forward an idea of integrating the automated generation of unique e-mail aliases into account registration workflows. 
While this practice has its own merits and interest from the surveyed participants, our results hint at limitations of this approach, too. 
One-time emails may function as double-edged swords, as they may make it harder for victims to learn about a breach notification. 
In addition, their utility is uncertain in the light of a permanent shipping address with the ensuing---perceived or real---risk of burglary or physical extortion in the worst case. 
In this regard, future research may explore workable design options for account registration and order checkout processes which would anticipate potential data leakages and adverse effects thereof.

\subsection{Implications for Breach Research}\label{section:implresearch}
In terms of methodology, our study has taught us some unanticipated lessons.
As we started recruitment, we were surprised to see that some victims made quite some efforts to convince themselves of the authenticity of our invitation as well as of the researcher's identity. 
For example, they contacted other university staff members through various channels.
While this is generally a positive sign of high risk awareness, our ethical analysis missed these activities as a form of additional (though very minor) cost on victims, researchers, and researchers' colleagues. 
It is unimaginable what would have happened had we used a non-existing persona as sender; an idea we contemplated to protect the lead researcher, but dismissed eventually.
Related to this, at the time of applying for IRB approval, we were unaware of the vendor's post-breach initiative to crowd-source reports of suspicious phishing campaigns from the community. 
Therefore, our study might have inadvertently created additional work for the vendor's customer support agents processing those notices.
We tried to inquire the cost after the study has ended to inform future ethics assessments, but we did not receive an estimate.
Likewise, we were unaware of the vendor's take-down efforts, which might have increased the risk of blacklisting and related collateral damage to other users of our university's infrastructure.
In hindsight, this appears obvious given the size of their business and the availability of professional services. 
We encourage other researchers to take these contingent actions into account in their ethics analyses.

Finally, we observe that many forms of experienced harm repeat across case studies of breaches~\cite{romanosky2014empirical,thomas2017data,peng2019happens}. 
Therefore, data breach research could benefit from the development of standard survey questions and response scales.  
This includes the refinement of measures of harm, an incredibly difficult concept with high relevance also in the field of breach litigation~\cite{kilovaty2021psychological}.  
More standardized measurement instruments would allow for more harmonized reporting of empirical results, improve comparability between cases, and pave the way for meaningful quantitative meta-analyses in the future.
\subsection{Limitations}\label{section:limitations}
Our work has a number of limitations inherent to survey-based cybercrime studies. 
It is evident that findings from a sample of volunteers among victims affected by a high-profile breach should be interpreted with caution. 
Known sources of coverage error include the receipt of the invitation, which is correlated with the e-mail provider's spam filtering practices.
Moreover, we could not reach victims who implemented certain security precautions, such as using one-time e-mail addresses.
Each individual decision to participate in our survey may be correlated with victim experience, for example the salience of perceived or actual harm; as well as victim's aptitude to report on sensitive topics in English.
Other causes of non-response are victims' subjective expectations towards receiving personalized invitations and reminders of pending survey invitations, which we refrained from sending for ethical considerations. 
All this explains the low response rate~\cite{cook2000meta}. 
Yet the sample size appears decent given that we have confirmed victims of a high-profile breach (cf.~Footnote~\ref{fn:power}). 
Other studies interview thousands in order to obtain a sample of a few hundred victims of everyday cybercrime~\cite{Riek2018}.

Turning to item reliability and validity, our survey instrument lacks effective mechanisms to verify and prevent inaccurate reporting. 
With self-reported statements, we cannot rule out response errors, strategic responses in hope for redress, and social desirability bias~\cite{zou2018ve}. 
Also, recall bias and the capability of causal reasoning may vary between individuals. 
While post-breach attacks can occur with some delay, individuals' ability to accurately recollect past events tends to degrade. 
This makes it tricky to optimize the timing of surveys which collect data on reciprocal causal effects. 
We hope that our choice of one year gave appropriate time for the influencing factors and effect variables to manifest, while important episodes are still in memory.
A longitudinal design could in principle overcome some of these shortcomings, however it is even harder to implement in an ethical manner. 
Moreover, recollections of effects of this data breach may be confounded by other breaches and attacks.
Finally, the survey instrument may be prone to unknown ceiling effects resulting from the chosen question wording and response scales~\cite{chyung2020evidence}. 
We had to make some compromises given the scarcity of standardized and validated item batteries in our domain.
Closing this gap---certainly not on high-profile cases at first---could catalyze the development of breach research.
\section{Conclusion}\label{section:conclusion}

We document one of the first empirical case studies on high-profile data breaches affecting vendors of critical security products.
While in this case the security of the product itself was not affected, the disclosed sales information has enabled or facilitated new attack vectors.
At the same time, it enabled us, researchers, to ethically, legally, technically, and practically explore the practice of sampling an otherwise hard-to-reach victim population from leaked contact information for the purpose of breach research.
We have reported existential and (limited) quantitative evidence on the case, derived lessons learned, and demonstrated the feasibility of the sampling technique.
The effort is tremendous and researchers have to exercise a lot of patience before getting hold of data or results.
Time will tell if our approach becomes more commonly accepted and eventually easier to pursue. 
What we can state with more certainty is that the security community is well-advised in learning from high-profile breaches.

\section*{Acknowledgments}
The authors are grateful to the anonymous reviewers and the paper shepherd for their valuable critical comments on the manuscript. We also thank J\'{e}r\'{e}mie Glossi for his technical support and valuable contributions to the study, our colleagues Nicole Krismer-Stern and Paulina Jo Pesch for legal consultation, Leonid Risteski for his help in producing Figure~\ref{fig:timeline}, and Daniel W. Woods for very thoughtful comments on a draft of this paper.

\bibliographystyle{plainnat}
\bibliography{references.bib}

\appendix
\section{E-Mail Invite} \label{app:emailinvite}
\textbf{Subject line:} Follow-up on the Ledger data breach\\

Hello,

I am \ifanonym{\dummyresearcher}{ }, a security researcher at the \ifanonym{\dummyuni}{ }. I am contacting you regarding the data breach of Ledger's customer database last year. I sincerely regret that personal information had been compromised and leaked to the public. Our team of security researchers would like to follow up on this incident and estimate the scope of effects and potential harm it caused to the victims. With this initiative, we strive to collect valuable first-hand insights from you. This will help us to make recommendations for better protection and response in order to avoid similar cases in the future. If you support this research initiative, I would like to invite you to participate in our {\bf anonymous online survey} available at \ifanonym{\dummylink}{\url{ }}.

Please note that your participation in this study is {\bf voluntary}. You are free to not participate at all or to withdraw from the survey at any time. Completing the survey will take about {\bf 25 minutes} of your time. Your anonymous responses will be analyzed and used by the members of our team for research purposes only. The results will be published in aggregate form in academic journals and conferences.

This research study has been approved by the Research Board for Ethical Issues of the \ifanonym{\dummyuni}{ } under the principles of freedom of research and beneficial societal impact. We, the researchers in charge, aim to collect opinions, experiences, and sentiments of the users affected by the data breach in order to derive policy recommendations and managerial implications for firms offering secure storage devices. The data controller of any personal data related to this study is the \ifanonym{\dummyuni}{ }. Processing takes place under the General Data Protection Regulation (GDPR) of the European Union.

The only piece of personal data processed is your {\bf e-mail address}. The processing is justified by the public interest of the conducted research (Article 6, p. 1(e) GDPR). Your e-mail address will be processed until the end of the collection of 500 reliable response sets or until the expiration of 6 months since the start of the research study on November, 8th 2021. Your e-mail address is processed only for the purpose of inviting you to participate in this study and will not be shared with other parties. It will be permanently deleted after the end of the term of processing (on May 8th, 2022) or immediately after your written request to \ifanonym{\dummycontact}{\texttt{ }}.

We take the following security measures to protect your e-mail address:

\begin{itemize}
	\item[--] Your e-mail address is stored in an encrypted form on an external USB flash drive, which is kept in a secure physical vault. The access to the vault and the drive is limited to authorized staff of the research group.
	\item[--] This invitation is sent from an e-mail account purposefully created and used for this study only. The account as well as all the entire communication will be permanently deleted after completion of the data collection phase.
\end{itemize}

In matters related to data processing, you may also contact the data protection coordinator of the \ifanonym{\dummyuni}{ } at \ifanonym{\dummydataprotec}{\texttt{ }}. You have the right to lodge a complaint with the data protection supervisory authority if you believe that the processing of your e-mail address for the purpose of this research study infringes the provisions of the GDPR.

The anonymous online survey is accessible at \ifanonym{\dummylink}{\url{ }}. I sincerely thank you for your collaboration.

Sincerely,

\ifanonym{\dummyresearcher}{ }

\ifanonym{\dummyuni}{}

\section{Static Webpage} \label{app:staticpage}
\textbf{Welcome to Our Survey on the Ledger Data Breach}\\

\textbf{\small Thank you for your interest in our study!}\\

\textbf{Access to the Questionnaire}\\

\textcolor{blue}{Please click here to access} the anonymous online survey. Completing the survey will take about 25 minutes of your time.

The questionnaire is hosted by \textcolor{blue}{Qualtrics}, a professional company offering survey tools to researchers. The company is bound by law and contracts to follow our high data protection standards. We are using this static webpage, hosted by the researchers, between the invitation link in the e-mail you received and the actual questionnaire to reassure all respondents that the links are not personalized. \textbf{This research respects your privacy}.\\

\textbf{Contacts}\\

If you have any questions or concerns about the project, please do not hesitate to contact us at  \ifanonym{\dummycontact}{databreach-contact-informatik@uibk.ac.at}.

In matters related to data processing, you may also contact the data protection coordinator of the  \ifanonym{\dummyuni}{University of Innsbruck} at \ifanonym{\dummydataprotec}{\texttt{datenschutzkoordination@uibk.ac.at}}. You also have the right to lodge a complaint with the data protection supervisory authority if you believe that the processing of your e-mail address for the purpose of this research study infringes the provisions of the GDPR.

\section{Questionnaire\ifquestionnaire{}{\protect\footnote{The complete questionnaire is available in an arXiv version of this paper.}}} \label{app:questionnaire}
Welcome and thank you for supporting our research initiative!

{\bf This research respects your privacy}. The survey does not collect personal data in the form of your name or contact details and we ask you not to provide this information in your open text responses. The survey is intended to be anonymous. If you disclose information concerning harm to yourself or another person within this survey, the researchers will not be able to take any action.

{\bf What is the purpose of this research?} We use this questionnaire to learn more about the effects and potential harm resulting from the Ledger data breach in 2020, which may have impacted you. We would love to hear your personal story of this incident. These insights will help us estimate the aggregate consequences of this data breach as well as understand your attitudes, security behaviors, and coping strategies.

{\bf Who are we?} We are researchers of the \ifanonym{\dummyuni}{}. We conduct this research independently of the Ledger company. This research is funded from the research budget of the \ifanonym{\dummyuni}{public university}.

{\bf Your participation is voluntary.} You do not have to participate in this survey, and you can leave it at any point. Completing the survey will take about 25 minutes of your time.

{\bf Your responses are anonymous}. We assure that we will remove any information which would make you identifiable. Your anonymous responses will be analyzed and reported in aggregate form and will be used for research purposes only.

{\bf There are potential benefits from participating.} You have the opportunity to share your story and thoughts about the data breach with us. Your valuable input will contribute to advance scientific research, improve companies’ security practices, and inform policy making in this domain.

{\bf There are potential risks from participating.} Responding to our questions may remind you of your experience as a victim of crime and any pain you may have experienced.

{\bf How to contact us?} If you have any questions or would like to get further information about this study, you may contact the principal investigator \ifanonym{\dummyresearcher}{} \ifanonym{\dummycontact}{(\texttt{})}.

{\bf Consent.} By clicking "Yes", you confirm to be over 18 years of age and consent to us collecting information about your attitudes, experiences, and demographic information. Detailed information regarding this project and your participation has been explained to you.

\ifquestionnaire{
\begin{enumerate}[label=\textbf{\arabic*}., itemsep=5pt, parsep=0pt]

	\item {\bf Do you consent to participating in this study?} \newline\singlechoice
	\begin{enumerate}
		\item Yes, I do consent.
		\item No, I do not consent.
	\end{enumerate}
\end{enumerate}

\begin{enumerate}[resume,label=\textbf{\arabic*}., itemsep=5pt, parsep=0pt]
	\item {\bf Ledger, the producer of hardware wallets, has experienced a data breach. When and how did you learn about this?}	\opentext
	
	\item {\bf Have you been victim of another breach since January 2020?} \singlechoice
	\begin{enumerate}
		\item Yes, with certainty.
		\item No.
		\item I don't know.
	\end{enumerate}

	\item {\bf For which purpose have you purchased a hardware wallet from Ledger?} \multiplechoice
	\begin{enumerate}
		\item To store crypto-assets for personal use
		\item As a giveaway or a gift to others		
		\item For resales		
		\item For a business purpose		
		\item For another purpose (please specify).		 
		\item I don't know / I don't recall.		
		\item I have not purchased the hardware wallet.		
	\end{enumerate}

	\item {\bf Do you use the purchased hardware wallet at the moment?} \singlechoice
	\begin{enumerate}
		\item Yes, I use it at the moment.		
		\item No, I discontinued to use it before the data breach occurred.
		\item No, I discontinued to use it after the data breach occurred.		
		\item No, I have never used it. 		
		\item I don’t know / I am not sure.		
		\item I prefer not to answer.		
	\end{enumerate}

	\item {\bf Do (or did) you use hardware wallets of other brands?
	} \singlechoice
	\begin{enumerate}
		\item Yes, I use one or more at the moment.		
		\item Yes, I used one or more in the past.		
		\item No, but I am aware of other brands and alternative products.		
		\item No, I am not aware of other brands and alternative products.		
		\item I don’t know / I am not sure.
		\item I prefer not to answer.
	\end{enumerate}

	\item {\bf Please list the main factors that have influenced your decision to choose a hardware wallet among other alternatives for storing crypto-assets.} \opentext
	
	\item {\bf Which factors influenced your decision to buy a hardware wallet from the present company specifically?}\newline\multiplechoice
	\begin{enumerate}
		\item Official registration and location		
		\item Terms of use		
		\item Amount of personal data I need to provide		
		\item Ease of opening an account		
		\item Technical security feature
		\item Reputation of the company		
		\item Supervision by a renowned authority		
		\item Supported crypto-assets and services		
		\item Recommendations from friends and family		
		\item Recommendation from members of the community		
		\item Lack of choice or alternatives		
		\item Other (please specify)		
		\item I don’t know / I don’t recall.
	\end{enumerate}

	\item {\bf Apart from hardware wallets, which alternative wallets do you use at the moment to store crypto-assets? For more precise definitions, please click on the information button.} \multiplechoice
	\begin{enumerate}
		\item Software wallet
		\item Paper wallet kept offline 	
		\item Other cryptographic storage device
		\item Brain wallet
		\item Multi-signature wallet	
		\item Mobile wallet	 		
		\item Cloud/online wallet
		\item Exchange
		\item Other (please specify)
		\item	None
		\item	I don’t know / I am not sure.	
	\end{enumerate}

	\item {\bf Think back to a time when you first purchased your hardware wallet and remember your original expectations about this product. Please specify to which extent you agree with the following statements.		
	} (Strongly disagree, Disagree, Neither agree or disagree, Agree, Strongly agree, I don't know.)
	\begin{enumerate}
		\item	I expected the hardware wallet to be secure for the storage of crypto assets.
		\item	I expected the hardware wallet to be a good choice for my personal needs.
		\item	I expected the present company and its products to be trustworthy.	
	\end{enumerate}

	\item \label{original_attitude_co}{\bf Try to recall your attitudes before you learned about the data breach. Please rate your overall attitude towards the company at the time before you learned about the data breach.		
	} (Very negative, Negative, Neither positive nor negative, Positive, Very positive, I don't know.)

	\item {\bf Please rate your overall attitude towards the product at the time before you learned about the data breach.		
	} (Very negative, Negative, Neither positive nor negative, Positive, Very positive, I don't know.)

	\item \textit{(if \ref{original_attitude_co} "Very negative" or "Negative":)} {\bf Why did you have a negative attitude towards Ledger before you learned about the data breach?} \multiplechoice
	\begin{enumerate}
		\item	I disapproved the company’s way of doing business.
		\item	I disapproved the company’s leadership team.
		\item	I had a bad experience with their customer service.
		\item	I had concerns about security of their products.
		\item	I had concerns about security of their firmware.
		\item	I found the price was too high for the value delivered.
		\item	The company had a negative publicity in the media.
		\item	The company had a negative publicity in the community of crypto asset users.
		\item	Other (please specify)
		\item	I don’t know / I don’t recall.	
	\end{enumerate}

	\item {\bf Now turn to the situation as of today. Please rate your overall attitude towards the company after you learned about the data breach.} (Very negative, Negative, Neither positive nor negative, Positive, Very positive, I don't know.)
	
	\item {\bf Please rate your overall attitude towards the product after you learned about the data breach.} (Very negative, Negative, Neither positive nor negative, Positive, Very positive, I don't know.)
	
	\item {\bf Think about your actual experience with your hardware wallet. To what extent do you agree with the following statements about the product?} (Strongly disagree, Disagree, Neither agree or disagree, Agree, Strongly agree, I don't know.)
	\begin{enumerate}
		\item	This hardware wallet is more advanced than any other product of this type.
		\item	This hardware wallet features advanced security guarantees.
		\item	There is no alternative hardware wallet of the same overall quality.
		\item	Looking back, the decision to purchase a hardware wallet from this company was right.
		\item	The hardware wallet fully meets my personal requirements.
		\item	The hardware wallet offers good value for money.
		\item	The hardware wallet functions trouble-free.	
	\end{enumerate}

	\item {\bf How satisfied are you with the hardware wallet based on all your user experience?} (Very dissatisfied, Somewhat dissatisfied, Neither satisfied nor dissatisfied, Somewhat satisfied, Very satisfied, I don't know.)
	
	\item {\bf To what extent has your hardware wallet met your expectations?} (Much less than expected, Less than expected, As expected, More than expected, Much more than expected, I don't know.)
	
	\item {\bf How well do you think the Ledger's hardware wallet compares to similar products of competitors?} (Very inferior, Somewhat inferior, Equivalent, Somewhat superior, Very superior, I don't know.)
\end{enumerate}

\begin{enumerate}[resume, label=\textbf{\arabic*}., itemsep=1pt, parsep=0pt, before=\setlength{\baselineskip}{2mm}] 
	
	\item {\bf How did you first find about this data breach incident?} \singlechoice
		\begin{enumerate}
			\item	From a notification e-mail sent out by the company 
			\item	From an official blog post published on the company’s website
			\item	From the company’s newsletter
			\item	From social media 
			\item	From my peers or colleagues
			\item	From TV news or newspapers
			\item	From online media sources
			\item	From dubious activities on my accounts
			\item	From my crypto wallet being emptied 
			\item	From other sources (please specify)
			\item	I don’t know / I don’t recall.
		\end{enumerate}
	
	\item {\bf How concerned are you that this incident may affect:} (Not at all concerned, Slightly concerned, Somewhat concerned, Moderately concerned, Extremely concerned, I don't know.)
	\begin{enumerate}
		\item	the security of your crypto assets
		\item	the security of your online accounts
		\item	your personal identity
		\item	your physical security
		\item   your mental health
	\end{enumerate}

	\item {\bf In your opinion, how likely is it that your crypto-assets stored on the purchased hardware wallet would be stolen as a consequence of this incident.} (Very unlikely, Unlikely, Neither likely nor unlikely, Likely, Very likely, I don't know.)

\end{enumerate}

\begin{enumerate}[resume, label=\textbf{\arabic*}., itemsep=5pt, parsep=0pt]  
	\item  \label{loss}{\bf Did you lose crypto-assets since July 2020?} \newline\singlechoice
	\begin{enumerate}
		\item	Yes, due to a theft
		\item	Yes, due to a scam
		\item	Yes, due to my own mistake
		\item	No
		\item	I don’t know. / I am not sure.
		\item	I prefer not to answer
	\end{enumerate}

	\item  \textit{(if \ref{loss} "Yes":)} {\bf How much of crypto-assets did you lose? Please specify the order of magnitude (or the exact amount) and the unit of the crypto-asset (e.g. BTC).} (Amount / Unit)
	
	\item  \textit{(if \ref{loss} "Yes":)} {\bf How likely is it that the loss of crypto-assets was caused by the data breach incident?} (Very unlikely, Unlikely, Neither likely nor unlikely, Likely, Very likely, I don't know.)

	\item \textit{(if \ref{loss} "Yes":)} {\bf From which wallet(s) did you lose crypto-assets?} \multiplechoice
	\begin{enumerate}
		\item	Ledger’s hardware wallet
		\item	Other hardware wallet
		\item	Software wallet
		\item	Paper wallet
		\item	Other cryptographic storage device (e.g., a card)
		\item	Brain wallet
		\item	Multi-signature wallet
		\item	Mobile wallet
		\item	Cloud/online wallet
		\item	Exchange
		\item	Other (please specify)
		\item	I don’t know / I am not sure.
		\item	I prefer not to answer
	\end{enumerate}

	\item \label{harm}{\bf Since July 2020, what harm(s) have you experienced?} \multiplechoice
	\begin{enumerate}
		\item	I had no harm resulting from this data breach at all.
		\item	I received marketing e-mails, SMS messages or phone calls.
		\item	I received scam or phishing e-mails, SMS messages or phone calls.
		\item	I received threatening (ransom) emails, SMS, or phone calls. 
		\item	I received password reset requests for accounts registered with the leaked e-mail address.
		\item	I received a scam letter mailed to my postal address.
		\item	I received a package mailed to my postal address with a fake hardware wallet.
		\item	Someone tried to register a different SIM with my phone number.
		\item	Someone managed to register a different SIM with my phone number.
		\item	Someone tried to take over the control over my online account(s).
		\item	Someone managed to take over the control over my online account(s). 
		\item	Someone tried to register a new online account with my e-mail address.
		\item	Someone managed to register a new online account with my e-mail address.
		\item	I have experienced some forms of cyberbullying.
		\item	I have received physical security threats.
		\item	Other (please specify the details)
	\end{enumerate}

	\item {\bf You might have been harmed by this incident in different ways. However, a loss that occurred after July 2020 was not necessarily caused by the data breach or in relation to it. Please indicate your confidence level that the harm you experienced was caused by this specific breach incident:} \multiplechoice (Very unlikely, Unlikely, Neither likely nor unlikely, Likely, Very likely, I don't know.) \textit{(items selected in \ref{harm})} 

	\item {\bf How did you respond to the data breach and its consequences?} \multiplechoice
	\begin{enumerate}
		\item	I spent time resolving issues on my own.
		\item	I followed the security advice provided by the company.
		\item	I needed to get in touch with the police.
		\item	I sought for advice or consultation from a professional security expert or company.
		\item	I sought for advice or clarification from the company.
		\item	I sought for advice or consultation from members of the crypto community.
		\item	I sought for advice or consultation from my peers and colleagues.
		\item	I sought for advice or consultation from other online sources.
		\item	I followed the discussion of this data breach on the company’s website.
		\item	I followed the discussion of this data breach incident on online platforms.
		\item	I shared my experience with the data breach incident on online platforms.
		\item	Other (please specify)
		\item	There was no need for me to respond to this data breach.
		\item	I don't know / I don't recall.
	\end{enumerate}

	\item {\bf Did you take any legal actions in relation to this incident?} \multiplechoice
	\begin{enumerate}
		\item	I am part of / I support a lawsuit against Ledger in front of a local court.
		\item	I am part of / I support a lawsuit against Ledger in front of a French court.
		\item	I contacted my local data regulator.
		\item	I contacted the French data regulator (CNIL).
		\item	I contacted the French government cybersecurity agency (ANSSI).
		\item	No, only the French courts are competent to receive a complaint.
		\item	No, I do not see a need for legal actions against the company.
		\item	No, I am not protected by the GPDR as a non-EU customer.
		\item	No, for other reasons.
	\end{enumerate}

	\item {\bf How do you keep yourself informed about the data breach?} \multiplechoice
	\begin{enumerate}
		\item	I do not keep myself informed about the company’s communication.
		\item	From notification e-mails send out by the company
		\item	From the company’s newsletter
		\item	From the blog posts published on the company’s website
		\item	From the company’s YouTube channel
		\item	From reddit.com
		\item	From other social media (Twitter, Facebook…)
		\item	From my peers or colleagues
		\item	From TV news or newspapers
		\item	From online media sources
		\item	From a lawyer
		\item	From other sources (please specify)
		\item	I don’t know.
	\end{enumerate}

	\item {\bf To what extent do you agree with the following statements concerning Ledger’s response to this incident?} (Strongly disagree, Disagree, Neither agree or disagree, Agree, Strongly agree, I don't know.)
	\begin{enumerate}
		\item	The company publicly acknowledged the full responsibility for the data breach.
		\item	The company sincerely apologized for the data breach.
		\item	The company apologized for the inconvenience or concern this data breach may have caused.
		\item	The company “sugar coated” the data breach in its official statement.
		\item	The company downplayed the severity of the data breach.
		\item	The company downplayed the implications of the data breach.
		\item	The company was transparent on taking security measures to prevent data breaches from reoccurring.
		\item	The company was transparent on taking proactive measures to reduce potential harm and damage caused by the data breach.
		\item	The company was conclusive in taking security and privacy of its customers seriously.
		\item	The company provided useful security advice on how to reduce the potential harm and damage caused by the data breach.
		\item	The company provided a comprehensive amount of information about the causes of the data breach.
		\item	The company provided a comprehensive amount of information about the potential effects of the data breach.		
	\end{enumerate}

	\item {\bf Companies take a variety of actions to resolve incidents where personal information is lost or stolen. Please rate the following actions in terms of how much they would affect your satisfaction with a company’s response following a loss/theft of your personal information.} (Would not improve my satisfaction at all, Would slightly improve my satisfaction, Would somewhat improve my satisfaction, Would moderately improve my satisfaction, Would greatly improve my satisfaction, I don't know.)
	\begin{enumerate}
		\item	Apologize to you
		\item	Notify you immediately 
		\item	Take measures to ensure that a similar breach cannot occur in the future
		\item	Donate money to nonprofit organizations that promote cybersecurity
		\item	Provide financial compensation to you for your inconvenience
		\item	Follow up and take necessary measures to ensure that lost data cannot be misused
		\item	Offer recommendation or security advice on how to protect yourself		
	\end{enumerate}

	\item \label{co_direct_contact}{\bf  Have you been in direct contact with the company regarding the consequences of the data breach?} \newline\singlechoice
	\begin{enumerate}
		\item	Yes, I contacted Ledger.
		\item	Yes, Ledger contacted me
		\item	No.
		\item	I don’t know / I am not sure.
		\item	I prefer not to answer.
	\end{enumerate}

	\item \textit{(if \ref{co_direct_contact} "Yes":)}\label{complaint_satisf} {\bf How would you rate the handling of your complaint/request by Ledger?} (Very poorly, Poorly, Neutral, Well, Very well, I don't know.)
	
	\item \textit{(if \ref{complaint_satisf} is "Very poorly" or "Poorly":)} {\bf What specifically did not go well in your interaction with Ledger? } \newline\opentext
	
	\item {\bf To what extent do you agree with the following statements referring to your general expectations and satisfaction with a company's response to a data breach?} (Strongly disagree, Disagree, Neither agree or disagree, Agree, Strongly agree, I don't know.)
	\begin{enumerate}
		\item	I expect monetary compensation when my personal data is leaked.
		\item	I expect non-monetary compensation (such as a discount or free premium services) when my personal data is leaked. 
		\item	I expect an apology from a company following up on a data breach.
		\item	I expect that a company shows remorse to its customers in case of a data breach.
		\item	I find that an apology is a reasonable response from a company in case of a data breach.
		\item	Overall, I am satisfied with how Ledger responded to this data breach.
		\item	The response of Ledger fully meets my expectations.
		\item	Looking back, I perceive the company’s response as a good experience.		
	\end{enumerate}

	\item {\bf To what extent do you agree with the following statements about liabilities? } (Strongly disagree, Disagree, Neither agree or disagree, Agree, Strongly agree, I don't know.)
	\begin{enumerate}
		\item	Ledger is fully responsible for this data breach.
		\item	Customers who lost crypto assets in relation to this data breach are fully responsible for their own losses. 		
	\end{enumerate}

	\item {\bf Did you take any of the following actions in response to this incident?} \multiplechoice
	\begin{enumerate}
		\item	I discontinued to use the provided e-mail address or switched to a new one.
		\item	I discontinued to use the leaked phone number.
		\item	I discontinued to use the hardware wallet device.
		\item	I discontinued my relationship with Ledger.
		\item	I requested Ledger to delete my personal data.
		\item	Other (please specify):
		\item	None of the above.
		\item	I don’t know.
	\end{enumerate}

	\item {\bf Below you find a non-exhaustive list of security practices one may adopt in response to this incident. Did you personally take any of the following actions?} \newline\multiplechoice
	\begin{enumerate}
		\item	I relocated my hardware wallet to another, more secure physical place.
		\item	I relocated my recovery seed phrase to another, more secure physical place.
		\item	I changed/updated a password for my account(s) related to crypto assets.
		\item	I changed/updated a password for my e-mail account.
		\item	I notified others who may also have been affected by the breach.
		\item	I started (or significantly extended the use of) using a password manager.
		\item	I started using (or significantly extended the use of) multi-factor authentication for my online accounts.
		\item	I started using (or significantly extended the use of) the hardware-based multi-factor authentication (e.g, Yubikey, Titan)
		\item	I started using (or significantly extended the use of) authenticator apps.
		\item	I transferred my crypto assets stored on an exchange to another crypto wallet.
		\item	I transferred my crypto assets stored on the hardware wallet to another crypto wallet. 
		\item	I asked other companies to delete my personal data or online accounts.
		\item	I improved my home security (e.g., by installing a visual security alarm system, updating locks, investing in a safe etc.). 
		\item	I contacted my telecom provider to prevent SIM swapping attacks.
		\item	Other (please specify)
		\item	I did not take any actions.
		\item	I don't know.
	\end{enumerate}

	\item {\bf If you could place a euro or dollar value on the amount of inconvenience, distress and annoyance this incident may have caused, what would it be? When answering this question, please ignore financial losses of crypto assets (if you had any) and do not forget to specify your currency of choice.} (Amount / Currency)
	
	\item {\bf In your opinion, what would be a reasonable amount of compensation offered by the company for inconvenience, distress and annoyance this incident may have caused? When answering this question, please ignore financial losses of crypto assets (if you had any) and do not forget to specify your currency of choice.} (Amount / Currency)
	
	\item {\bf  How much time did it take in total to resolve issues related to this incident?} \singlechoice
	\begin{enumerate}
		\item	Less than 1 day
		\item	Between 1 and 3 days
		\item	Between 3 and 7 days
		\item	Between 1 week and 1 month
		\item	More than one month
		\item	I don’t know.
	\end{enumerate}

	\item \label{prec_curr}{\bf Below you find a non-exhaustive list of security practices one may adopt in response to this incident. Did you personally take any of the following actions?} (Yes, I started doing it before the breach; Yes, I started doing it after the breach; No)
	\begin{enumerate}
		\item	Using PO boxes or private mailboxes for the delivery of my online orders.
		\item	Using a one-time “burner” e-mail address for each online order.
		\item	Using a dedicated e-mail address for business related to crypto assets.
		\item	Using a dedicated SIM card for business related to crypto assets.
		\item	Using postal forward services for the delivery of my online orders
		\item	Using a decoy crypto asset storage device in case of a physical extortion
		\item	Not using my real name when ordering online
		\item	Not providing my real phone number when ordering online
	\end{enumerate}

	\item {\bf How likely is that you would take the following precaution actions in response to this incident?} (Very unlikely, Unlikely, Neither likely nor unlikely, Likely, Very likely) \textit{(items from \ref{prec_curr} answered with "No")}

	\item {\bf Please specify how often you undertake the following security practices.} (Rarely, Occasionally, Regularly, I don't know.)
	\begin{enumerate}
		\item	I store my crypto assets in a reputable online wallet or exchange.
		\item	I back up my crypto wallet.
		\item	I generate multiple backups of my crypto wallet.
		\item	I encrypt backups for additional security.
		\item	I keep my hardware wallet and its backup key separately.
		\item	I take care of a physical security of a seed required to re-store my access to the crypto wallet.   
		\item	I enable a multiple-factor authentication for my online account(s).
		\item	I disconnect from the Internet before creating private keys.
		\item	I use a multi-signature crypto wallet out of security concerns.
		\item	I store private keys differently depending on the purpose and amount of crypto assets.
	\end{enumerate}

	\item {\bf Now think of a device you most often use to access your crypto-assets. Please specify how often you undertake the following security practices in regards to this device.} (Never, Sometimes, Always, I don't know.)
	\begin{enumerate}
		\item	The device has a unique password.
		\item	The device requires my fingerprint, facial recognition, a PIN or passcode to be activated.
		\item	The device is equipped with the latest malware protection.
		\item	The device is kept in a physically secured location.
		\item	The device is used by someone else.
		\item	The device is connected to the Internet.
	\end{enumerate}

	\item {\bf Who has control over private keys for the majority of your crypto assets (in terms of value)?} \singlechoice
	\begin{enumerate}
		\item	I myself and no one else.
		\item	Me and other trusted person(s)
		\item	Service (e.g., an exchange or online wallet)
		\item	I don’t know / I am not sure.
		\item	I prefer not to answer.
	\end{enumerate}

	\item {\bf Taking into account all your customer experience and effects of the data breach into account, please specify to which extent you agree with the following statements.} (Strongly disagree, Disagree, Neither agree or disagree, Agree, Strongly agree, I don't know.)
	\begin{enumerate}
		\item	I will continue to use the Ledger hardware wallet.
		\item	I will not change to another hardware wallet’s provider after the data breach.
		\item	In the future, I intend to use further products or offers of Ledger.
		\item	I consider myself to be a loyal customer of Ledger.
		\item	I would recommend the hardware wallet device of Ledger to those who seek my advice about such matters.
		\item	I would encourage others to use products of Ledger.
		\item	Hardware wallet devices of Ledger would be my first choice for the secure storage of crypto-assets.
		\item	I speak positively about Ledger to other people.
		\item	I would post positive messages about Ledger on online platforms.
		\item	I intend to continue to do business with Ledger.
		\item	It would cost me too much to switch to another hardware wallet provider.
		\item	It would take me a great deal of time and effort to get used to a new provider.
		\item	In general, it would be a hassle switching to another provider.		
	\end{enumerate}

\end{enumerate}

\begin{enumerate}[resume, label=\textbf{\arabic*}., itemsep=5pt, parsep=0pt]  
	\item {\bf How many years of experience with crypto assets do you have?} \singlechoice
	\begin{enumerate}
		\item	Less than 1 year
		\item	Between 1 and 2 years
		\item	Between 3 and 4 years
		\item	Between 5 and 6 years
		\item	More than 6 years
	\end{enumerate}
	
	\item {\bf How much, in terms of the market value, are you currently holding in crypto assets?} \singlechoice
	\begin{enumerate}
		\item	Less than USD 1,000
		\item	Between USD 1,000 and USD 5,000
		\item	Between USD 5,000 and USD 10,000
		\item	Between USD 10,000 and USD 100,000
		\item	Between USD 100,000 and USD 1,000,000
		\item	More than USD 1,000,000
		\item	I prefer not to answer.
		\item	I don't know.
	\end{enumerate}

	\item {\bf How old are you?} \singlechoice
	\begin{enumerate}
		\item	24 or younger
		\item	Between 25 and 34
		\item	Between 35 and 44
		\item	Between 45 and 54
		\item	Between 55 and 64
		\item	65 or older
		\item	I prefer not to answer.
	\end{enumerate}
	
	\item {\bf In which country do you currently reside?} \newline \singlechoice 
	
	\textit{(list of all countries)}
	
	\item {\bf What is your gender?} \singlechoice
	\begin{enumerate}
		\item	Man
		\item	Woman
		\item	Non-binary/third gender
		\item	I prefer to self-describe as:
		\item	I prefer not to answer.
	\end{enumerate}

	\item {\bf What is your current occupation?} \singlechoice
	\begin{enumerate}
		\item	Student
		\item	Skilled manual worker
		\item	Employed position in a service job
		\item	Self-employed/freelancer
		\item	Unemployed or temporarily not working
		\item	Retired or unable to work through illness
		\item	Employed professional
		\item	Other
		\item	I prefer not to answer.
	\end{enumerate}

	\item {\bf What is the highest degree you have received or the highest level of education you have completed?} \newline\singlechoice
	\begin{enumerate}
		\item	Less than high school
		\item	High school incomplete
		\item	High school graduate (or an equivalent)
		\item	College or associate degree
		\item	Bachelor’s degree
		\item	Master’s degree
		\item	Doctoral degree
		\item	Other postgraduate or professional degree (please specify)
		\item	I prefer not to answer
	\end{enumerate}
\end{enumerate}
}{}
\end{document}